\documentclass[a4paper,11pt]{article}
\usepackage{amsmath}
\usepackage{amssymb}

\usepackage{times}

\usepackage{enumerate}
\usepackage{enumitem}

\usepackage{amsmath}
\usepackage{amssymb}
\usepackage[dvips]{graphicx}

\usepackage{fullpage}
\usepackage{appendix}
\usepackage{hyperref}
\usepackage{scrextend}
\usepackage{fixmath}
\usepackage{float}
\usepackage{authblk}
\usepackage{color}

\newcommand{\remove}[1]{}

\newtheorem{claim}{Claim}

\newtheorem{proposition}{Proposition}

\newtheorem{corollary}{Corollary}

\newtheorem{definition}{Definition}

\newtheorem{lemma}{Lemma}

\newtheorem{theorem}{Theorem}

\newtheorem{remark}{Remark}

\newcommand\G{\mathbold{G}}

\newcommand\MaxDegEvent{\mathbold{A}}

\newcommand{\WeightPathA}{ {\cal Z}}
\newcommand{\WeightPathB}{ {\cal H}}

\newenvironment{proof}{\noindent{\bf Proof\@:}}{\hfill $\Diamond$
 \\}

\newenvironment{lemmaproof}[1]{\noindent{\bf Proof of Lemma #1:\@}}{\hfill $\diamond$}
\newenvironment{claimproof}[1]{\noindent{\bf Proof of Claim #1:\@}}{\hfill $\diamond$}

\title{MCMC sampling colourings and independent sets of\\ $G(n,d/n)$ near uniqueness threshold}
\author{Charilaos Efthymiou\footnote{Partially supported by DFG grant EF 103/1-1.}\\
 Goethe University, Frankfurt 60054, Germarny\\
{\tt efthymiou@gmail.com}}

\begin{document}

\maketitle				

\begin{abstract}

\noindent
Sampling from {\em Gibbs distribution} is a central problem in computer science.
In this work we focus on the {\em $k$-colouring model} and the 
{\em hard-core model} with fugacity $\lambda$ when the underlying graph is an instance of 
Erd\H{o}s-R\'enyi random graph $G(n,d/n)$, where $d$ is fixed.

We use the {\em Markov Chain Monte Carlo} method for sampling from the aforementioned 
distributions. In particular, we consider {\em Glauber (block) dynamics}.
We provide significantly  improved  bounds for {\em rapid mixing} in terms 
of the number of colours and the fugacity for the corresponding models. 
For both models the bounds we get are only within small constant 
factors from the optimal ones,  conjectured  by the statistical physicists.

We use {\em Path Coupling} to show rapid mixing. For $k$ and $\lambda$ in the range of 
our interest the technical challenge is to cope with the high degree vertices, 
i.e. vertices of degree much larger than the expected degree $d$. The natural approach 
to this problem is to consider block updates rather than single vertex updates for the Markov chain. 
Using appropriately defined blocks the effect of high degree vertices somehow diminishes.
However, devising such a construction of blocks is a {\em  non trivial} task. 

We develop for a first time a weighting schema for the paths of the underlying graph.
Vertices which belong to ``light'' paths, only, can be placed at the boundaries 
of the blocks. 
This gives rise to simple structured blocks, i.e. trees with at most one extra edge.
\end{abstract}


\section{Introduction}

\noindent{\em
Let $\G=G(n,d/n)$ denote the random graph on the vertex set $V(\G)=\{1,\ldots,n\}$ where each edge
appears independently with probability $d/n$, for a sufficiently large fixed number $d>0$.

We say that an event occurs {\em with high probability (w.h.p.)} if the probability 
of the event to occur tends to 1 as $n\to \infty$.
}
\\

\noindent
Sampling from {\em Gibbs distributions} is a central problem in computer science as well as
in statistical physics. 
In this work we focus on colourings and independent 
sets when the underlying graph is an instance of Erd\H{o}s-R\'enyi random graph
$\G=G(n,d/n)$, where $d$ is a `sufficiently large' number but remains bounded as $n\to \infty$.
	
Given the graph instance $\G$, the focus is on two different kinds of Gibbs distribution. The first one
is the {\em colouring model}, i.e. the uniform distribution over the $k$-colourings 
of $\G$. The second one is the {\em hard-core model} with 
{\em fugacity} $\lambda$, i.e. the independent set $\sigma$ is selected with 
probability proportional to $\lambda^{|\sigma|}$. The parameters of interest 
are the number of colours $k$ and the fugacity $\lambda$.

The most powerful and, somehow, the most natural algorithms for sampling are based on the Markov Chain Monte 
Carlo (MCMC) method. Consider some graph $G$ and some target distribution $\mu$, e.g., the $k$-colouring model.
The algorithm simulates a Markov chain whose configuration space is the set of $k$-colourings of $G$, while the equilibrium 
distribution of the chain is the Gibbs distribution. Starting from some initial configuration, the algorithms simulates the Markov 
chain for a certain number of transitions. Then, it outputs the configuration of the chain. The objective is that the algorithm 
simulates the chain for sufficiently many steps  that the distribution of the output configuration is close to the equilibrium 
distribution of the chain.

A natural measure of efficiency for a MCMC algorithm is the speed that the underlying Markov chain converges to the
equilibrium. That is, the faster the chain mixes the faster the algorithm achieves the approximation guarantees we require 
for the output. The main technical challenge  is to establish that the corresponding Markov
chain exhibits  {\em rapid mixing} 
(see \cite{FrVigSurvey,jerrumA,jerrumB}) in the range of parameters of interest.

The Markov chains we consider here is discrete time {\em Glauber dynamics}. 
 At each time step the chain updates a small block of vertices, i.e. we have block dynamics. The aim is to show
rapid mixing for $k$ as small as possible and $\lambda$ as large as possible for the corresponding models.

For Glauber dynamics to mix fast, typically, the bounds for both $k$ and $\lambda$ are expressed in terms of the 
{\em maximum degree} of the underlying graph. Examples of such bounds are  \cite{COLPaperA,RCLocalySparse,RCLattice,RCPlanar,MartinelliTrees,MolloyLagreGirth,RCGeneralBound}
for colouring and \cite{ISPaperA,ISPaperB,MyFOCS16,MartinelliTrees,ISPaperC} for independent sets. From this perspective, 
what makes the case of $\G$ interesting is the relatively big fluctuations in the degree of its vertices. To be more specific, 
w.h.p. the vast majority of vertices in $\G$ are of degree close to $d$, while the maximum degree is as huge as 
$\Theta\left(\frac{\ln n}{\ln\ln n}\right)$. In such a situation, it is natural to expect that the rapid mixing bounds for both
 $k$, $\lambda$ depend on the {\em expected degree} $d$, rather than maximum degree.

Sophisticated but mathematically non rigorous arguments from statistical physics (e.g., in \cite{PNAS}) support this picture. 
They suggest that w.h.p. over the instances of $\G$ the Glauber (block) dynamics on $k$-colouring has rapid mixing for any 
$k>d$. Furthermore, for $k<d$ the chain is expected to be non-ergodic and  weaker notions of convergence hold.
Similarly, as far as the hard-core model is concerned the   conjecture is that we have rapid mixing as long as $\lambda \leq
\frac{(d-1)^{d-1}}{(d-2)^{d}}\approx \frac{e}{d}$.

So far, the best bounds for Monte Carlo sampling, for both colouring model and hard-core model on $\G$, 
appear  in \cite{mossel-colouring-gnp} (which improved on 
\cite{old-GnpSampling}). The authors in \cite{mossel-colouring-gnp} provide, for the first time, rapid mixing bounds
for $k$ and $\lambda$ which depend on the expected degree $d$. That is, w.h.p. over $\G$ there are functions $f(d)$ and
$h(d)$ such that Glauber dynamics has rapid mixing for $k$-colourings and hard-core for $k\geq f(d)$ and $\lambda\leq h(d)$,
respectively\footnote{Even though these functions are not given explicitly it is conceivable from the analysis that it holds that 
$f(d)\geq d^c$ and $h(d)\leq d^{-c'}$ for fixed $c,c'>0$.}. However, the values for $k$ and $\lambda$ that are allowed there are
some orders of magnitude off the conjectured bounds. Here we improve on these bounds siginificantly. We show that w.h.p. over 
the underlying graph $\G$ we have rapid mixing for $k\geq \frac{11}{2}d$ and for $\lambda\leq \frac{1-\epsilon}{2}\frac{1}{d}$. 
That is, we approach the conjectured bounds for rapid mixing only within small constant factors.

We use the well-known {\em Path Coupling} technique \cite{PathCouplingPaper} to show rapid mixing.  For
$k$ and $\lambda$ in the range of  interest the technical challenge in applying Path Coupling is to cope with 
the high degree vertices, i.e. vertices of degree  larger than $d$. The natural approach is to consider block
updates rather than single vertex updates for the Markov chain. What motivates the use of blocks is the observation
that the effect of high degree vertices diminishes when these vertices are away from the boundary of their block. The
improvements on the rapid mixing of Glauber dynamics we get in terms of $k$ and $\lambda$ rely on proposing
a good set of blocks for the dynamics. 
Devising such block construction is a {\em highly complex} task. 

For defining the blocks, we introduce, for a first time, a weighting schema for the paths of
the underlying graph. Specifically, we use a weighting schema to assign weight to 
each path in $\G$.   These weights allow to distinguish which vertices can be used for the boundaries of the blocks. 
These are every vertex   that has all the paths emanating from it of  small weight.
We call such vertices {\em break-points}. 
 Typically,  there is a plethora of break-points in $\G$. This allows creating small, simple structured blocks.

Subsequently to our work, the use of weighting of paths found application for sampling-colouring of $G(n,d/n)$ 
using a non-MCMC approach. The work in \cite{FPRAUS} use  essentially the same weighting schema as the one
we introduce here, to  propose  a non Monte Carlo FPRAUS for approximate sample of colourings of $k\geq 3d$.

\paragraph{Notation}
Given some graph $G$, we let $V(G)$ and $E(G)$ denote the vertex sets and the edge set, respectively.
We use small letters of the greek alphabet to indicate colourings or independent sets, e.g., $\sigma$, $\tau$. Also,
by $\sigma(v)$ we indicate the assignment of the vertex $v$ under the configuration $\sigma$. 
For a vertex set $B$ we define $\partial B$ to be  the {\em (outer) boundary} of $B$, i.e. 
\[
\partial B=\{v\in V(G)\setminus B \ | \   \exists w\in B \ s.t.\ \{v,w\}\in E(G) \}.
\]

\subsection{The Algorithm}\label{sec:ModelsChains}

\remove{
We consider  distributions which are defined w.r.t. some  graph $G=(V,E)$ and a set of {\em spins} $C$. 
These are  the {\em colouring model} and the {\em hard core model}.
Each one of them is a distribution over the {\em configuration space} $\Omega \subseteq C^V$.  Sometimes,
we refer to either of the two distributions as {\em Gibbs distribution}.

\paragraph{Colouring Model.}
Given a graph $G$ and a sufficiently large integer $k$, the {\em colouring model} is the uniform  distribution over the
proper $k$-colourings of $G$.

\paragraph{Hard Core Model.} Given a graph $G$ and $\lambda>0$, the hard core model with {\em fugacity} 
$\lambda$ is a distribution over the independent sets of $G$.  Each independent set $\sigma$ is assigned probability
which is proportional to  $\lambda^{|\sigma|}$, where $|\sigma|$ is the cardinality of $\sigma$.  \\

\noindent
}

The algorithm we propose for 
  has the following, general, form:

\paragraph{Algorithm:} The input is a graph $G$, a specification for the target distribution $\mu$ and ${\tt err}\in [0,1]$. 
By specification of $\mu$, we mean the parameters $k, \lambda$ for the colouring and the hard-core model,
respectively. The quantity ${\tt err}$ expresses the maximum distance between the distribution of the  configuration
at the output of the algorithm and $\mu$.
\\ \vspace{-.2cm}

\noindent
${\tt SetUp:}$ The algorithm partitions the set of vertices $V(G)$ into  disjoint small blocks. For the instances of graphs
we  consider,  each block is a tree with
at most one extra edge. Let ${\cal B}$ denote the set of blocks.

Using the set of blocks {\cal B} we define an appropriate $(X_t)_{t\geq 0}$ whose limit distribution
is $\mu$. The algorithm simulates this chain  and outputs the configuration of the chain after $T=T({\tt err})$  transitions.
\\ \vspace{-.2cm}

\noindent
${\tt Markov\; Chain:}$ The chain is a discrete time one. The initial configurations $X_0$ is an {\em arbitrary} one. 
\begin{itemize}
	\item Let  $X_t$ be the  configuration  at time $t$.  Then $X_{t+1}$, the configuration at time 
	$t+1$ is acquired as follows: 
	
	\begin{itemize}
		\item Chooses uniformly at random (u.a.r.) a block $B\in {\cal B}$ 
		\item For every vertex $u\notin B$ set $X_{t+1}(u)=X_{t}(u)$
		\item Set $X_{t+1}(B)$ according to the distibrution $\mu$ conditional  $X_{t+1}(V\backslash B)$.
	\end{itemize}

\end{itemize}

\noindent
The chain that is used by the algorithm, above, is the  well  known  {\em Glauber block dynamics}.
It is an easy exercise to show that if the chain has a single limit distribution, i.e. the chain is ergodic, then
this  distribution  is $\mu$ (e.g., see \cite{PeresMixingBook}).

For   a measure of distance between two distributions $\nu_{a}, \nu_{b}$ on some space $\cal S$, 
we use  the notion of {\em total variation distance}  $|| \nu_{a}-\nu_{b}||$ which is defined as follows:
\begin{displaymath}
|| \nu_{a}-\nu_{b}||=\max_{ A \subseteq {\cal S}} | \nu_{a}(A)-\nu_{b}(A)|.
\end{displaymath}

\noindent
The main result of this work is in the following theorem.

\begin{theorem}\label{thrm:MAIN}
Let $\epsilon>0$ be fixed and let $d$ be sufficiently large. On input $\G=G(n,d/n)$ and ${\tt err}\in [0,1]$ for the above algorithm 
the following is true:
\begin{description}

\item[colouring model:] W.h.p. over the instances $\G$ and for $k\geq \frac{11}{2}d$ the algorithm returns a 
$k$-colouring of $\G$ distributed within total variation distance ${\tt err}$ from Gibbs distribution. The time 
complexity of the algorithm is $\ln \left(1/{\tt err} \right)\times O(n^c)$, for a fixed $c>0$.

\item[hard-core model:] W.h.p. over the instances $\G$ and for $\lambda\leq\frac{1-\epsilon}{2d}$ the algorithm returns
an independent set of $\G$ distributed within total variation distance ${\tt err}$ from Gibbs distribution. The time complexity 
of the algorithm is $\ln \left({1}/{{\tt err}}\right)\times O(n^{c'})$, for a fixed $c'>0$.

\end{description}
\end{theorem}

\noindent
One of the most challenging tasks in proving Theorem \ref{thrm:MAIN} is to show that the bock dynamics
mixes rapidly for the range of the parameters which we consider. 
We use {\em mixing time}, $\tau_{mix}$, as a measure of the speed of convergence of Markov chains. The mixing time is defined 
as the number of transitions needed in order to guarantee that the chain starting from an arbitrary  configuration, is within total 
variation distance $1/e$ from the stationary distribution (see \cite{PeresMixingBook}). For a Markov chain  we say 
that it is   {\em rapidly mixing} if $\tau_{mix}$ is polynomial in $n$, the number of vertices of the underlying graph.

Theorem \ref{thrm:MAIN} follows as a corollary from a sequence of results that we present in the following section.

\subsection{Analysis of the algorithm}\label{sec:result}

Mainly, the efficiency of the algorithm depends on two factors. The first one is how
efficiently can the algorithm simulate the block dynamics. The second one is how
fast does the dynamics converge. In this section we are dealing with these two issues.

We start by considering the simulation of the block dynamics by the algorithm.
This problem is reduced to studying the following questions:

\begin{enumerate}
	\item What is the  number of steps required by the algorithm to compute the set of blocks ${\cal B}$
	\item What is the time required by the algorithm to implement a single transition of 
	the block dynamics 
	\item What is the number of step required to get an initial configuration of the chain.
\end{enumerate}

\noindent
As far as the time complexity of creating the set of block $\cal B$  is regarded   we provide the following 
theorem.
\begin{theorem}\label{thrm:FastBlockCreation}
With probability $1-o(1)$ over the instances of $\G$ the following is true: 
\begin{enumerate}[ label=(\alph*) ]
  	\item $\G$ admits a partition of its vertex set into an appropriate set of blocks {\cal B}. 
	Each block $B\in {\cal B}$ is a tree with at most one extra edge.
	\item There is a small fixed $s>0$ such that ${\cal B}$  can be computed in time $O(n^s)$. 
\end{enumerate}
\end{theorem}
The proof of Theorem \ref{thrm:FastBlockCreation} appears in Section \ref{sec:BlckCreation}.

As far as   the second issue is regarded, we note the following:
The fact that the blocks in $\cal B$ are trees with at most one extra edge implies that the update of the block dynamics can be 
implemented efficiently. In particular we have the following result whose proof is somehow standard. 
\begin{corollary}\label{cor:UpdateComplexity}
If each block in ${\cal B}$ is a tree with at most one extra edge, then the number of steps for each update is $O(n^s)$, for small 
fixed $s>0$. 
\end{corollary} 
For the sake of completeness we provide the proof of Corollary \ref{cor:UpdateComplexity}  in  Section \ref{sec:cor:UpdateComplexity}.

Next we consider the third issue, i.e. the time complexity of getting an initial configuration
for the block dynamics.  The following remarks imply that, for typical instances of $\G$, the algorithm can compute in 
polynomial time an initial configuration of the block-dynamics, for both hard-core model and
colouring model .

\begin{description}
\item [Hard-core model:]  We can trivially  consider the empty independent 
set as the initial configuration of the block dynamics. That is,  the time complexity 
of getting an initial configuration is $O(1)$.

\item [Colouring model:]  We can get an initial  configuration for the block dynamics
by using the algorithm  proposed in \cite{grimmett}.  This is  a (greedy), {\em polynomial time} algorithm
 which  $k$-colours typical instances  of $\G$ for any $k\geq (1+c)\ln d/d$ and any  $c>0$. 
\end{description}

\noindent
From the above we conclude that for typical instances of $\G$, the algorithm can simulate the underlying 
block-dynamics for both colouring model and hard-core model.

As far as the convergence of the Markov chains is regarded, this reduces to two questions.
The first one is whether the block-dynamics converges to a unique distribution. 
For this,  we need to show that the Markov chains  satisfy a set of   conditions which are 
known as {\em ergodicity}. This guarantees that  indeed there is a limit distribution.
The second question, which is the most challenging one, is what is the mixing time of the
block-dynamics.

The second question is dealt in the following two theorems. The first one is for the colouring model and
the second for the hard-core.

\begin{theorem}\label{thrm:Col-MainResult}
Let $d$ and $k$ be as in Theorem \ref{thrm:MAIN}. With probability $1-o(1)$ over the graph instances $\G$ the following is true: 

$\G$ admits a partition of its vertex set  into set of blocks ${\cal B}$ as described in Theorem \ref{thrm:FastBlockCreation}. 
Let   $(X_t)_{t\geq 0}$ be the discrete time block dynamics for the $k$-colouring model on $\G$ with set of blocks $\cal B$.  
Then, $(X_t)_{t\geq 0}$  is ergodic and the mixing time is $O(n\ln n)$.
\end{theorem}

\begin{theorem}\label{thrm:IS-MainResult}
Let $d$ and $\lambda$ be as in Theorem \ref{thrm:MAIN}. With probability $1-o(1)$ over the graph instances $\G$ the following is 
true: 

$\G$ admits a partition of its vertex set  into set of blocks ${\cal B}$ as described in Theorem \ref{thrm:FastBlockCreation}. Let   $(X_t)_{t\geq 0}$ be the discrete time block dynamics for the hard-core model on $\G$ with set of blocks $\cal B$ 
and fugacity $\lambda$.  
Then, $(X_t)_{t\geq 0}$  is ergodic and the mixing time is $O(n\ln n)$.
\end{theorem}

\noindent
Then,  Theorem \ref{thrm:MAIN} follows as a corollary from  the above results  and the following observation. 
\begin{remark}
It is standard  to show that the number of transitions required to get within error ${\tt err}$ from the stationary distribution is 
$T({\tt err})\leq \ln\left( {1}/{{\tt err}}\right)\times \tau_{mix}$.
\end{remark}

\section{Rapid mixing - Proof techniques}\label{sec:technique} 

Before getting into the details about the speed of convergence  let us introduce some concepts. 
For the sake of concreteness we focus on $k$-colourings.
Given some graph $G$ and two $k$-colourings $\sigma, \tau$, we let $\sigma\oplus \tau$
be the set of disagreements, i.e. the set of vertices $w$ such that $\sigma(w)\neq \tau(w)$. Also, we let the Hamming distance $H(\sigma,\tau)$  be defined as 
\[
H(\sigma,\tau)=\sum_{w\in V(G)} \mathbf{1}\{w\in \sigma\oplus \tau\},
\]
where $\mathbf{1}\{w\in \sigma\oplus \tau\}$ is an indicator variable of the event that $w\in \sigma\oplus \tau$.

We show rapid mixing by using the well-known  {Path Coupling}  technique.
First, we consider  the single site Glauber dynamics on the $k$  colourings of  a graph $G$,  for some large  
$k>0$.  Let $(X_t)_{t\geq 0}$, $(Y_{t})_{t\geq 0}$ be two copies of the Glauber dynamics. For some $t$, assume 
that  $X_t,Y_t$ are such that $H(X_t,Y_t)=1$.

We have rapid mixing if there is a coupling such that 
\begin{equation}\label{eq:PathCouplingCritirion}
\mathbb{E}[H(X_{t+1},Y_{t+1})|X_t,Y_t]\leq 1-\Omega\left(n^{-1}\right).
\end{equation}

\noindent
To study the technique further, assume that the underlying graph $G$ is of maximum degree $\Delta$, 
while  $k>\Delta$. Also, let $X_t\oplus Y_t=\{w\}$.

It is natural to use a coupling that updates the same vertex in both copies. Then, the cases that matter are only those 
where the coupling chooses to update either the disagreeing vertex $w$  or one of the neighbours of $w$. 
If  vertex  $w$ is updated, then we can couple the two copies  such that $X_{t+1}=Y_{t+1}$. The probability the dynamics chooses 
to update $w$ is $1/n$. 
On the other hand, if the dynamics updates $u$, a  neighbour of $w$, then it is possible that $X_{t+1}(u)\neq Y_{t+1}(u)$, i.e.
we  get an extra  disagreement. This is due to the  disagreement at $w$.
In the worst case, the update of $u$ causes a new disagreement with probability at most  $\frac{1}{k-\Delta}$. 
Since the  disagreeing vertex $w$ has at most $\Delta$  neighbours, the probability of having an extra disagreement between 
$X_{t+1}$ and $Y_{t+1}$  is at most  $\frac{\Delta}{n}\frac{1}{k-\Delta}$. Then, we get that
\begin{equation}\label{eq:PathCouplingIntroductoryVerySimple}
\mathbb{E}[H(X_{t+1},Y_{t+1}) \ |\  X_t,Y_t]\leq 1 -\frac{1}{n}+ \frac{\Delta}{n}\frac{1}{k-\Delta}.
\end{equation}
The above implies for  $k\geq 2\Delta+1$,  the criterion in (\ref{eq:PathCouplingCritirion})  is satisfied.

Typically,  $\G$ is of maximum degree $\Theta\left(\frac{\ln n}{\ln\ln n}\right)$. The above arguments imply that
the ``vanilla path coupling"  requires an unbounded number of colours. Otherwise, i.e. if $k$ is smaller than the maximum 
degree,  there is no control on the expected number of disagreements generated.  It turns out that we 
gain some  control over the expected number of disagreements by using (appropriate) {\em block updates} rather than 
single vertex ones.
In particular,  we need to have the  high degree vertices ``hidden'' inside the  blocks, i.e. no high degree vertex is close to the
boundary of its block.

\begin{remark}
The idea of using a block construction with the properties we describe above, was  first proposed in \cite{old-GnpSampling}. 
To a great extend it was also adapted in the subsequent work \cite{mossel-colouring-gnp}.
\end{remark}

\noindent
Consider two copies of the block dynamics, with set of blocks ${\cal B}$, i.e. $(X_t)_{t\geq 0}$ and
$(Y_t)_{t\geq 0}$. Assume that for some $t\geq 0$, we have $X_t$ and $Y_t$ such that $X_t\oplus Y_t=\{w\}$.
We couple one transition of the two copies.  The coupling chooses to update the same block in each copy.
 It turns out that the crucial case for  (\ref{eq:PathCouplingCritirion})  is when the outer boundary of 
$B$ is not the same for both chains, i.e. $w \in \partial  B$. 
In particular,  if $\Gamma$ is the set of block which have $w$ in their outer boundary, it holds that
\[
\mathbb{E}[H(X_{t+1},Y_{t+1})\ |\ X_t,Y_t]\leq 1 -\frac{1}{|{\cal B}|}+ \frac{1}{|{\cal B}|}\sum_{B\in \Gamma} 
\mathbb{E}[|X_{t+1}(B)\oplus Y_{t+1}(B)| \ |\ X_t,Y_t, \textrm{$B$ is updated}],
\]
i.e.  $|X_{t+1}(B)\oplus Y_{t+1}(B)|$ is the  number of extra {\em disagreements} generated inside $B$, given
that it is updated at time  $t+1$.

We use the well-known ``{\em disagreement percolation}'' coupling construction \cite{DisPerc} to bound the expected number
of disagreements in the block $B$. The disagreement at the boundary prohibits identical coupling of $X_{t+1}(B)$ and $Y_{t+1}(B)$.
The disagreement percolation assembles the coupling in a stepwise fashion moving away from $w$. Disagreements propagates into $B$  along paths from $w$. A disagreement at vertex $v \in B$ at distance $r$ from $w$ propagates to a neighbour $u$ at distance 
$r+1$ if  $X_{t+1}(u) \neq Y_{t+1}(u)$.

The disagreement percolation is {\em dominated} by an independent process such that each vertex $u \in B$ is disagreeing with probability
\begin{equation}\label{Def:DisagreementProbs}
\varrho(u)=\left\{
\begin{array}{lcl}
 \frac{2}{k-(1+\alpha)d} &\qquad&\textrm{if ${\tt degree}(v) \leq (1+\alpha)d$}\\
1 && \textrm{otherwise},
\end{array}
\right.
\end{equation}
where $\alpha>0$ is a  small constant.  The disagreement propagates over the path $L$ that start from $w$ with  probability at most 
$\prod_{u\in L\backslash\{w\}}\varrho(u)$. The expected number of disagreements after the update of $B$ is at most the expected number of paths of 
disagreements that start from $w$ and propagate  inside $B$.

Intuitively, the contribution of  high degree vertices, i.e. of degree larger than $(1+\alpha)d$, to the disagreements in $B$ is increased.  
If a high degree vertex is disagreeing, it has an increased number of  neighbours to propagate the disagreement. 
On the other hand, having $k=\Theta(d)$, the low degree vertices have relatively small probability of propagating the disagreement,
i.e. probability $O(1/d)$.
Having the large degree vertices at large distance from the boundary of $B$, makes the disagreement less probable to reach them.
For the block construction we propose, the high degree vertices are in such a long distance from the boundary of their block
that  the reduced probability of a disagreement reaching them  ```counter balances" their increased contribution 
to the disagreements in the block $B$.

For defining the set of blocks $\cal B$, we introduce a weighting schema as follows: Each vertex $u$
is assigned weight $W(u)$ such that
\begin{equation}
 W(u)=\left \{
\begin{array}{lcl}
(1+\gamma)^{-1} &\qquad& \textrm{if ${\tt degree}(u)\leq (1+\alpha)d$}\\ \vspace{-.3cm}
\label{eq:DefOfW(vi)}\\ 
d^c \  {\tt degree}(u) && \textrm{otherwise},
\end{array}
\right .
\end{equation}
for appropriate fixed numbers $\alpha,\gamma,c>0$, which we specify later.

Given the weights of the vertices, each block  $B\in {\cal B}$  satisfies the following property:
For every path $L$ between a vertex in $\partial B$, i.e. the outer  boundary of $B$,
and a high degree vertex inside $B$  it  holds that 
\[
 \prod_{u\in L}W(u)\leq 1.
\]

In the weighting schema, observe that the low degree vertices reduce the weight of the path $L$, while  the high degree vertices 
increase it. Restricting the weight of a path between a high degree vertex in the block $B$ and a boundary vertex, somehow, 
guarantees  that the high degree vertices are sufficiently far from the boundary. 
\begin{remark}\label{rmrk:LightPaths}
Since   large weight -high degree- vertices are rather rare in a path $L$ in $\G$, 
we expect  the weight of $L$ to  be  rather low.  
 Note that    $\Pr[W(u)>(1+\gamma)^{-1}]\leq \exp\left(-\alpha^3d \right)$, for all $u\in L$.
\end{remark}

\noindent
In  Section \ref{sec:thrm:CL-Tail} we prove the following tail bound on the weight of a path of length $\ell$ in $\G$. 
\begin{theorem}\label{thrm:CL-Tail}
Let  $\alpha\in (0,3/2)$, $0<\gamma<4, c>0$  and  $\delta>0$ be fixed numbers. Consider the graph
$\G=G(n,d/n)$, for sufficiently large $d$.   Let $P=v_1,\ldots, v_{\ell}$ be permutation of vertices of 
$\G$ such that  $\ell\leq 100\ln n$.
It holds that
\[
\Pr\left [\prod^{\ell}_{i=1}W(v_i)   \geq \delta \;|\; {\cal E}_{P} \right]\leq 2\exp\left[-d^{4/5}\left(\ell+\ln \delta\right) \right],
\]
where  ${\cal E}_{P}$ is the event that  $P$ forms  a path in $\G$.
\end{theorem}

\paragraph{Choosing  the appropriate $k$.}
The previous discussion does not make clear how do  we choose the parameters $\alpha,\gamma,c$ and especially
the number of colours $k$. This turns out to be a bit technical argument  and it is related on the details of
 bounding the expected  number of disagreements in a block.   
Assume  a single disagreeing vertex $w$ and a block  which  has the disagreement at its
outer boundary. 
Due to its construction, the block has only one vertex adjacent to the  disagreeing vertex $w$. Let us call this vertex 
$v$. The block  will be a tree with at least one extra edge. W.l.o.g. assume that the block is a tree\footnote{If the 
block was unicyclic then would have to  consider the tree of ``self avoiding walks".}. Let us call it $T$, while the root is
 $v$, the vertex next to disagreement.

Consider the update of $T$ with the disagreement at $w$. Then, the expected number of disagreements at 
$T$ is dominated by the independent process  on $T$ where  each vertex is disagreeing with the probability
specified in (\ref{Def:DisagreementProbs}). 
Let $R_i(T)$ denote the expected number of paths of disagreements in $T$ that connect the root and the vertices 
at level $i$  of $T$. 
For the rapid mixing condition (\ref{eq:PathCouplingCritirion}), it will suffice to show that 
\begin{eqnarray}\label{eq:SketchCondition4DisPaths}
R_i(T)\leq q(1-\theta)^i \qquad \textrm{for $i\geq 0$},
\end{eqnarray}
for appropriate $\theta<1$ and $q>0$.   
The condition in (\ref{eq:SketchCondition4DisPaths}) can be reduced to the  condition:  
For each subtree $T'$ rooted at a child of the root of $T$  it   should hold that
\begin{eqnarray}\label{eq:SketchCondition1994}
R_{i-1} (T')\leq \frac{q(1-\theta)^i}{ {\tt degree}(v) \  \varrho_v}, 
\end{eqnarray}
where $\varrho_v$ is the probability of disagreement for the root of $v$ in the independent 
process. Note that $R_{i-1}(T')$ is the expected number of paths of disagreement from the root of $T'$
to the vertices at level $i-1$ of $T'$.

We   apply the same argument  $i$ times. Then we  consider the subtrees $T''$ which are rooted at  level 
$i$ of $T$, i.e. each subtree $T''$ contains 
only vertices which are at distance  $j\geq i$ from the root of $T$. The paths of disagreement we consider 
in each $T''$ is  of length zero,  i.e. the probability of the root to be disagreeing.
More precisely the condition we get  is the following one: For each subtree $T''$ rooted at level
$i$ of $T$  it should hold that 
\begin{eqnarray}\label{eq:SketchCondition2002}
R_{0}(T'')\leq \frac{q(1-\theta)^i}{\prod_{x\in L }\left({\tt degree}(x)\  \varrho_x\right)},
\end{eqnarray}
where $L$ is the path in $T$ from $v$ to the root of $T''$.

If $u$ is the root of $T''$ in \eqref{eq:SketchCondition2002}, then we have that  $R_{0}(T'')\leq p_u$.  This observation and \eqref{eq:SketchCondition2002} imply the 
following condition: For each subtree $T''$ rooted at vertex $u$, at 
level $i$ of $T$,  it should hold that 
\begin{eqnarray}\label{eq:SketchCondition1978}
\frac{q(1-\theta)^i}{\prod_{x\in L }\left({\tt degree}(x) \  \varrho_x\right)}\geq \varrho_u.
\end{eqnarray}

\noindent
 The denominator in (\ref{eq:SketchCondition1978}), as a product of degrees of the vertices on the
 path $L$, is related to the weight of the path $L$.   
Somehow the way of defining the weights is related to whether \eqref{eq:SketchCondition1978} is true or not.
That is, given
$q, \theta$ we adjust the parameters $\alpha,\gamma,c$ and $k$ so that
 (\ref{eq:SketchCondition1978}) holds for every $u$  at level $i$ of $T$, for every $i\geq 0$. 
The usual configuration for the parameters is that we choose some small $\delta>0$
and set  $\gamma<\alpha\leq \delta$, while $c=10$ and $k \geq \frac{11}{2}d$.

\begin{remark}
We follow exactly the same approach to show rapid mixing for the hard core model.  
\end{remark}

\section{Block Creation \& Proof of Theorem \ref{thrm:FastBlockCreation}}\label{sec:BlckCreation}

For the creation of block consider the weighting schema introduced in Section \ref{sec:technique}.
That is, we have $\G=G(n,d/n)$ and the parameters  $\alpha$, $\gamma$ and $c$. Each vertex
$w\in \G$ is assigned weight as specified in \eqref{eq:DefOfW(vi)}.
Given the weighting schema we introduce some useful notions.
\begin{definition}[Influence \& Break-Points]\label{def:influenece}
For a vertex $v$, let ${\cal P}(v)$ be  the set of all path  of length at most $\frac{\ln n}{d^{2/5}}$ that start from $v$.
We define the quantity $E(v)$, which we call  {\em influence},  as follows
\[
E(v)=\max_{L\in {\cal P}\left(v\right)}\left\{\prod_{ u \in L}W( u)\right\}.
\]
If $E(v)\leq 1$, then vertex $v$ is called {\em break-point}.
\end{definition}

\begin{definition}
The  path $L$ is called ``influence path" only  if non of its vertices is a break-point. 

If $w_1$ is a break-point, we define that there is only one influence path that starts from $w_1$, this is 
the trivial  path $L={w_1}$.
\end{definition}

To get some intuition, typically,  there is a very large number of break-points in $\G$. 
Somehow, this implies that we should expect to  have only short   influence paths.
As we will see later, this also implies that ${\cal B}$ contains very simple blocks.

The following pseudo-code describes how do we compute the set of blocks $\cal B$.
\\ \vspace{-.3cm}

\noindent
${\tt Block Creation}(\G, d, \alpha,\gamma,c)$\\ \vspace{-.7cm} \\
\rule{\textwidth}{1pt} \\ 
${\tt \;\;1:}$
\hspace*{.6cm}{\tt compute}  $\mathbf{\Sigma}$ the set of break-points\\
${\tt \;\;2:}$
\hspace*{.6cm}{\tt compute}  ${\cal C}$ the set of the set of all cycles of length at most  $4\frac{\ln n}{ (\ln d)^5}$ in $\G$\\
${\tt \;\;3:}$
\hspace*{.6cm}{\tt set} ${\cal B}$ to be empty\\
${\tt \;\;4:}$
\hspace*{.6cm}{\tt for-each} cycle $C\in {\cal C}$ do\\
${\tt \;\;5:}$
\hspace*{1.2cm}define block $B_c$ that contain the following vertices\\
${\tt \;\;6:}$
\hspace*{1.8cm}every $v$ in the cycle $C$ and  in $\partial C$\\
${\tt \;\;7:}$
\hspace*{1.8cm}{every $w$ for which there is an influence path from $w$ to a vertex in  $\partial C$}\\
${\tt \;\;8:}$
\hspace*{1.2cm}add  $B_c$ into $\cal B$\\
${\tt \;\;9:}$
\hspace*{.6cm}{\tt end-for} \\
${\tt 10:}$ 
\hspace*{.6cm}{\tt while} there is  a vertex $w$ whose block is not specified yet {\tt do}\\
${\tt  11:}$ 
\hspace*{1.2cm}define block $B_w$ that contain the following vertices\\
${\tt  12:}$ 
\hspace*{1.8cm}vertex $w$ \\
${\tt  13:}$ 
\hspace*{1.8cm}every $u$ that is reachable from $w$  through an influence path \\
${\tt  14:}$ 
\hspace*{1.2cm} add  $B_w$ into $\cal B$ \\
${\tt  15:}$ 
\hspace*{.6cm}{\tt end-while} \\
${\tt  16:}$  
\hspace*{.6cm}{\tt return} ${\cal B}$\\ \vspace{-.7cm} \\
\rule{\textwidth}{1pt} \\

\noindent
Typically,   $\G$ the distance between any two  cycles  in ${\cal C}$ is large. Since we don't expect to have long influence paths,  
there are no two cycles in $\cal C$ which are connected through an influence path.  The short influence paths also yield that
the rest of the blocks are trees. That is,   the blocks with the extra edge are exactly those which have a cycle from ${\cal C}$.

Note that each break-point is a block by itself.  One the other hand,  each multi-vertex block is created  such that its 
outer boundary  consists only of break-points.   This argument also implies that the blocks in $\cal B$ are {\em vertex disjoint}.

The reader may have observed that the definition of the break point considers only the weight
of paths with length at most $\frac{\ln n}{d^{2/5}}$.  We are going to show that typically $\G$ has the 
following property:  if a vertex $u$ has no heavy path in ${\cal P}(u)$, then it has no heavy  path at all. 
That is, for each $B\in {\cal B}$ and every $w\in \partial B$   there is no path $L$ from $w$ to a high degree 
vertex $u\in B$ such that $\prod_{v \in L}W(v)>1$.

\subsection{Proof of Theorem \ref{thrm:FastBlockCreation}}

For proving part (a) in  Theorem \ref{thrm:FastBlockCreation},  first we show that the influence paths that are considered 
in the construction of $\cal B$ are short.  Then, using this result we show that indeed  $\cal B$ consists of blocks that are 
trees with at most one extra edge.

In the statement of the following theorem, we call {\em elementary}  every path $L=w_1, \ldots, w_{\ell}$  such that there is no
other path $P$ \footnote{ i.e. $P$ is different than $L$} of length less than $10 \ln n/d^{9/10}$ which connects any two vertices in 
$L$.

\begin{theorem}\label{thrm:BreakablePaths}
Let  $\gamma,c>0$ and $\alpha\in (0,3/2)$. For large $d$, consider $\G=G(n,d/n)$. Let  $\mathbf{U}$ be the set of the 
{\em elementary} paths in $\G$ of  length  $\frac{\ln n}{ (\ln d)^{5} }$  that do not have any break-point.  It holds that
\[
\Pr[\mathbf{U}\neq \emptyset] \leq 4n^{(-\frac{1}{2}\ln d+2)}. 
\]
\end{theorem}
The proof of Theorem \ref{thrm:BreakablePaths} appears in Section   \ref{sec:thrm:BreakablePaths}.

Using Theorem \ref{thrm:BreakablePaths} it is direct to get the following  lemma,
whose proof appears in Section \ref{sec:lemma:SimpleBlock}.

\begin{lemma}\label{lemma:SimpleBlock}
Let $\gamma,c>0$ and $\alpha\in (0,3/2)$. For large $d$ consider $\G=G(n,d/n)$.
With probability at least $1-10n^{-3/4}$ over the instances $\G$, ${\cal B}$ contains only
blocks which are trees with one extra edge. 
\end{lemma}

Lemma \ref{lemma:SimpleBlock} proves that part (a) of Theorem \ref{thrm:FastBlockCreation} is  indeed true. As
 far as part (b) of Theorem \ref{thrm:FastBlockCreation} is regarded, it suffices to  use the following result.

\begin{lemma}\label{lemma:FastBlockCreation}
Let $\gamma,c>0$ and $\alpha\in (0,3/2)$. For large $d$, consider $\G=G(n,d/n)$. With probability at least $1-20n^{-3/4}$ over 
the instances of $\G$, the following is  true: 

There is a small fixed $s>0$ such that the block construction can be made  in time $O(n^s)$.
\end{lemma}
The proof of Lemma \ref{lemma:FastBlockCreation} appears in Section \ref{sec:lemma:FastBlockCreation}.

Theorem \ref{thrm:FastBlockCreation} follows.

\section{Convergence \& Rapid Mixing}

\subsection{Colouring Model - Proof of Theorem \ref{thrm:Col-MainResult}}\label{sec:thrm:Col-MainResult}

\paragraph{Ergodicity.}
For Theorem \ref{thrm:Col-MainResult} first we need to consider when the Markov chain  is {\em ergodic}. 
This allows to argue that the chain has  a unique limit distribution.   From \cite{old-GnpSampling} we have that 
the Glauber dynamics (and hence the  block dynamics) is ergodic with probability $1-o(1)$ over the instances 
$\G$ when $k\geq d+2$.

For the sake of completeness let us sketch the  proof for ergodicity in \cite{old-GnpSampling}.
It is shown that if a graph $G$ has no $t$-core\footnote{For some integer $r>0$ and a graph $G$, we say that $G$ 
has a $r$-core if it has a subgraph with minimum   degree $r$}, then for all $k\geq t+2$ the Glauber dynamics for 
$k$-colouring yields an ergodic Markov chain (Lemma 2  in \cite{old-GnpSampling}). Then the authors use the result  
in \cite{k-core}, which states that w.h.p. $\G$ has no $t$-core for $t\geq d$.
\paragraph{Rapid mixing.}
Given fixed $0<\gamma < \alpha\leq 10^{-2}$ and $c=10$ and large $d$,  let  ${\cal G}_{\chi}={\cal G}_{\chi}(n,d,\alpha,\gamma, c)$ 
be the family of graphs on  $n$ vertices such that $G\in {\cal G}_{\chi}$ if it has  the following properties:

\begin{enumerate}[ label=(\alph*) ]
\item  there is  no $r$-core, for $r \geq d$

\item  it is $k$-colourable for $k\geq d$

\item  Let ${\cal B}$ be the set of blocks constructed by ${\tt Block Creation}(G, d, \alpha,\gamma,c)$.
Then  each of the blocks in ${\cal B}$ is  a tree with at most one extra edge.

\item  For every $B\in {\cal B}$ and every $w\in \partial B$  there is no path $L$ from $w$ to a high degree 
vertex $u\in B$ such that $\prod_{v \in L}W(v)>1$ 
\end{enumerate}

\begin{lemma}\label{lemma:ExpDisGraphCond}
With probability $1-o(1)$ the graph $\G \in {\cal G}_{\chi}$.
\end{lemma}
The proof of Lemma \ref{lemma:ExpDisGraphCond} appears in Section \ref{sec:lemma:ExpDisGraphCond}.

Consider some $G\in {\cal G}_{\chi}$ and let $\cal B$ be the set of blocks  constructed by ${\tt Block Creation}
(G, d, \alpha,\gamma,c)$.  Consider also the block dynamics $(X_t)_{t\geq 0}$, over the $k$-colourings of $G$, 
for $k\geq \frac{11}{2}d$. The following theorem shows that $(X_t)_{t\geq 0}$ satisfies the path coupling condition.

\begin{theorem}\label{thrm:ExctDisagreement}
Let $0<\gamma < \alpha\leq 10^{-2}$ and $c=10$ and sufficiently large $d$. Let $G\in {\cal G}_{\chi}(n,d,\alpha,\gamma, c)$ 
and let $\cal B$ be the set of block created by  ${\tt Block Creation}(G, d, \alpha,\gamma,c)$. Let both $(X_t)_{t\geq 0}$, 
$(Y_t)_{t\geq 0}$  be the block dynamics over the $k$-colourings of $G$, with set of blocks ${\cal B}$.  For $k\geq \frac{11}{2}d$, the following is true:

Assume that for some $t\geq 0$ we have $H(X_t,Y_t)=1$.  Then, there is a coupling such that 
\begin{equation}\label{eq:ContrCond}
\mathbb{E}  [H(X_{t+1},Y_{t+1}) |X_t,Y_t ]\leq 1- ( 10 n)^{-1}. 
\end{equation}
\end{theorem}
The proof of Theorem \ref{thrm:ExctDisagreement} appears in Section \ref{sec:thrm:ExctDisagreement}.

Given the above results, Theorem \ref{thrm:Col-MainResult} follows as a corollary.

\subsection{Hard Core Model - Proof of Theorem \ref{thrm:IS-MainResult}}\label{sec:thrm:IS-MainResult}

\paragraph{Egodicity} 
Block dynamics for hard-core model is trivially ergodic. This follows from the observation that 
from every independent sets $\sigma$ there is a sequence of independent sets 
$\sigma=\sigma_0,\sigma_1, \ldots, \sigma_{\ell}$, for appropriate $\ell>0$, 
such that $\sigma_{\ell}$ is the {\em empty independent set},
while $\Pr[X_{t+1}=\sigma_{i} |  X_{t}=\sigma_{i-1}], \Pr[X_{t+1}=\sigma_{i-1} |  X_{t}=\sigma_{i}]>0$,
for $i=1, \ldots, \ell$.

\paragraph{Rapid mixing.}
For  $\epsilon>0$ as defined in the statement of Theorem \ref{thrm:IS-MainResult}, let $\epsilon_0=\min\{\epsilon, 1/100\}$.
Also let  $\alpha=\epsilon_0/3$, $\gamma=\epsilon^2_0$,  $c=10$ and large $d$. We define  
 ${\cal G}_{\alpha}={\cal G}_{\alpha}(n,d,\epsilon, c)$ to be the family of graphs on  $n$ vertices with the following properties: 

\begin{enumerate}[label=(\alph*)]

  \item  Let ${\cal B}$ be the set of blocks constructed by ${\tt Block Creation}(G, d, \alpha,\gamma,c)$.
Then  each of the blocks in ${\cal B}$ is  a tree with at most one extra edge.

\item  For each $B\in {\cal B}$ and every $w\in \partial B$  there is no path $L$ from $w$ to a high degree 
vertex $u\in B$ such that $\prod_{v \in L}W(v)>1$.

\end{enumerate}

\begin{corollary}\label{cor:IS-ExpDisGraphCond}
With probability $1-o(1)$ it holds that $\G\in {\cal G}_{\alpha}$.
\end{corollary}
The proof of Corollary \ref{cor:IS-ExpDisGraphCond} is very similar to
the proof Lemma \ref{lemma:ExpDisGraphCond}, so we omit it.

Consider some $G\in {\cal G}_{\alpha}$ and let $\cal B$  be the set of blocks constructed by ${\tt Block Creation}(G, d, \alpha,\gamma,c)$. Consider also the block dynamics $(X_t)_{t\geq 0}$ for the
hard core model with fugacity $\lambda=\frac{1-\epsilon}{2d}$.
The following theorem shows that $(X_t)_{t\geq 0}$ satisfies the path 
coupling condition.

\begin{theorem}\label{thrm:IS-ExctDisagreement}
Let $\epsilon>0$ be as defined in Theorem \ref{thrm:Col-MainResult} and $\epsilon_0=\min\{\epsilon, 10^{-2}\}$.
Also let  $\alpha=\epsilon_0/3$, $\gamma=\epsilon^2_0$,  $c=10$ and  sufficiently large $d>0$.
Consider  $G\in {\cal G}_{\alpha}$ and  let $\cal B$ be the set of block 
constructed by ${\tt Block Creation}(G, d, \alpha,\gamma,c)$.
Let both  $(X_t)_{t\geq 0}$,  $(Y_t)_{t\geq 0}$  be the block dynamics for the hard-core model  on  $G$, with set of blocks ${\cal B}$ and 
fugacity $\lambda=\frac{1-\epsilon}{2d}$.   Then,  the following is true:

Assume that for some $t\geq 0$ we have $H(X_t,Y_t)=1$. 
 There is a coupling such that 
\begin{equation}\label{eq:IS-ContrCond}
\mathbb{E}  [H(X_{t+1},Y_{t+1}) |X_t,Y_t ]\leq 1- {\epsilon_0} /({2n}).
\end{equation}
\end{theorem}
The proof of Theorem \ref{thrm:IS-ExctDisagreement} appears in Section
\ref{sec:thrm:IS-ExctDisagreement}.

Given the above results, Theorem \ref{thrm:IS-MainResult} follows as a corollary.

\section{Proof of Theorem \ref{thrm:ExctDisagreement}}\label{sec:thrm:ExctDisagreement}

The coupling is such that we choose the same block to update in both $(X_t)_{t\geq 0}$ and 
$(Y_t)_{t\geq 0}$. Let $B_t$ be the block whose colouring is chosen to be updated in the 
coupling at time $t$.

Assume that $X_t\oplus Y_t=\{w\}$, i.e. the two copies disagree only on vertex $w$.
Let $B_w$ be the block that contains  the disagreeing vertex $w$. Also,  let $\Gamma_w \subset {\cal B}\setminus \{B_w\}$ be 
the blocks that are adjacent to $w$, excluding $B_w$.
To simplify our further analysis we need to make the following observations:
\begin{description}
\item [Observation 1.] If $B_t=B_w$, then there is a coupling such that $H(X_{t+1}, Y_{t+1})=0$.
\item [Observation 2.] If $B_t \notin \Gamma_w\cup B_w$, then there is a coupling such that $H(X_{t+1}, Y_{t+1})=1$.
\item [Observation 3.] If $B_t=\{u\}\in \Gamma_w$, then $u$ is a break-point and $\Pr[X_{t+1}(B_t)\neq Y_{t+1}(B_t)]\leq \frac{1}{k-(1+\alpha)d}$.
\end{description}
Observation 1 and Observation 2 follow from  the fact that the assumption that $X_t\oplus Y_{t}=\{w\}$ 
implies that  $X_t(\partial B)=Y_t(\partial B)$, for every  $B \in {\cal B}\setminus \Gamma_w$.
If $B_t\in {\cal B}\setminus \Gamma_w$, then there is a coupling such that no new disagreements are created.
Furthermore,   if $B_t=B_w$, then  the disagreement vanish.

As far as Observation 3 is regarded, note  that since $u$ is a single-vertex block it is
a break-point and   ${\tt degree}(u)\leq (1+\alpha)d$. 
Consider the lists $L_X$ and $L_Y$ of available colours for $u$ that are induced by 
$X_{t+1}$ and $Y_{t+1}$, respectively.  Both lists are of size
at least $k-{\tt degree}(v)$.
Furthermore, these two lists differ in at most one colour.
Then there is a coupling for $X_{t+1}$ and $Y_{t+1}$ such that
$\Pr[X_{t+1}(u)\neq Y_{t+1}(u)]\leq ( {k-(1+\alpha)d})^{-1}$.

The following proposition generalizes Observation 3, in the sense that it considers
general blocks from $\Gamma_w$ and not necessarily single vertex ones.

\begin{proposition}\label{prop:ExpctNoDisBlock}
Let $\alpha,\gamma,c$ and $k$ be as in the statement of Theorem \ref{thrm:ExctDisagreement}. 
There is a coupling such that 
\[
\mathbb{E}[ H(X_{t+1}(B_t), Y_{t+1}(B_t) ) | X_t, Y_t, B_t\in \Gamma_w ] \leq {0.8688}/{d}.
\]
\end{proposition}
 The proof of Proposition \ref{prop:ExpctNoDisBlock} appears in Section \ref{sec:prop:ExpctNoDisBlock}.

In what follows,  we let $N=|{\cal B}|$, i.e. each $B\in {\cal B}$ is chosen with probability  $1/N$.
So as to prove \eqref{eq:ContrCond} we  consider a number of different cases for $B_w$ and the position
of $w$ inside $B_w$.\\ \vspace{-.3cm}

\noindent
{\tt Case 1:}
The  block $B_w$ is multi-vertex and $w$   is not adjacent to any vertex in $\partial B_w$.
Since $w$ is inside $B_w$ we have that $\Gamma_w=\emptyset$.
There is a coupling which has the following properties:
With probability  $1/N$  we have $B_t=B_w$.  Then, using Observation 1, we get that $H(X_{t+1},Y_{t+1})=0$.
Also, with  probability  $1-1/N$ we have $B_t\notin \Gamma_w\cup B_w$, since $\Gamma_w=\emptyset$. 
Then, using Observation 2 we get that  $H(X_{t+1},Y_{t+1})=1$
Combining all the above,  we get that 
\[
\mathbb{E}  [H(X_{t+1},Y_{t+1}) | X_t, Y_t ]=1-{1}/{N}. 
\]

\noindent
{\tt Case 2:}
The block $B_w$  is multi-vertex   and   $w$  is  adjacent to at least one vertex in $\partial B_w$.

Recalling the properties of the blocks in $\cal B$, our assumptions 
imply the following:
we have that ${\tt degree}(w)\leq (1+\alpha)d$.
Also,  $\Gamma_w\neq \emptyset$, while each  $B\in \Gamma_w$  is single vertex
block of a  break-point.  Note that a break-point
has degree at most $(1+\alpha)d$.

There is a coupling that has the following properties: With probability $1/N$ we have $B_t=B_w$ and, according to 
Observation 1, we have  $H(X_{t+1}, Y_{t+1})=0$.  Also, with probability $|\Gamma_w|/N$, we have that $B_t\in \Gamma_w$.
Since each $B\in \Gamma_w$ consists of a single break-point, from  Observation 3 we get that 
$\Pr[X_{t+1}(B_t)\neq Y_{t+1}(B_t)]\leq \frac{1}{k-(1+\alpha)d}$. Finally, if $B_t \notin \Gamma_w\cup B_w$, then, from 
Observation 2, we have $H(X_{t+1}, Y_{t+1})=1$.  With the above arguments, we get that
\[
\mathbb{E}  [H(X_{t+1},Y_{t+1}) | X_t, Y_t ]=1-\frac{1}{N}+\frac{1}{k-(1+\alpha)d} \frac{|\Gamma_w|}{N} \leq 1-\frac{3}{4N},
\]
since $|\Gamma_w|\leq {\tt degree}(w)\leq (1+\alpha)d$, $k\geq \frac{11}{2}d$ and $\alpha \le 10^{-2}$.
\\ \vspace{-.2cm}

\noindent
{\tt Case 3:}
$B_w$ is a single vertex block.   i.e. $w$ is a break-point.

With probability $1/N$ we have $B_t=B_w$. Then, according to Observation 1, we have   $H(X_{t+1},Y_{t+1})=0$. Also, with probability  
$|\Gamma_w| /N$ we have that $B_t\in \Gamma_w$.  Then using Proposition \ref{prop:ExpctNoDisBlock} we have that 
the number of  disagreements generated is  at most $0.8688/d$. Finally, if  $B_t\notin \Gamma_w \cup B_w$, we have that  
$H(X_{t+1},Y_{t+1})$=1, from Observation 2. Combining all the above together, we get that 
\[
\mathbb{E} [H(X_{t+1},Y_{t+1}) | X_t, Y_t] \leq 1-\frac{1}{N}+\frac{|\Gamma_w |}{N} \frac{0.8688}{d}  \leq 1-(10 n)^{-1}.
\]
since $\alpha \le 10^{-2}$ and $N\leq n$.
The theorem follows.

\subsection{Proof of Proposition \ref{prop:ExpctNoDisBlock}}\label{sec:prop:ExpctNoDisBlock}

Consider some $B\in \Gamma_w$. Recall that $X_t(w)\neq Y_t(w)$  and this is the only disagreement at
the boundary of $B$. Also, there is exactly one vertex in $B$ which is adjacent to $w$. Let us call this vertex
$r$. For each vertex $u\in B$ let $\mathbf{J}_u$ be an indicator which is 1 if ${\tt degree}(u)\leq (1+\alpha)d$ 
and $0$, otherwise.

We wish to couple $X_{t+1}(B)$ and $Y_{t+1}(B)$ as close as possible.
Identical coupling is  precluded for the whole block $B$ due to the disagreement at the boundary of $B$, recall that $w\in \partial B$.
The coupling we use for setting $X_{t+1}(B)$ and $Y_{t+1}(B)$ decides the colour assignment of each
vertex $u\in B$ at a time.  Every time we give priority to the vertices which are next to a disagreement,
i.e. we decide the colouring for these vertices first.  If there are vertices in $B$ whose colouring is not decided
but non of them is next to a disagreeing vertex, then it is easy to see that we can colour them using identical
coupling.

\begin{lemma}\label{lemma:DisProbBound}
Assume in the above process of deciding $X_{t+1}(B)$ and $Y_{t+1}(B)$, the process consider some
vertex $u\in B$.  There is a coupling  such that $u$ becomes disagreeing with probability  at most $ \rho_u$, where 
\begin{displaymath}
 \rho_u= \left( \frac{2\mathbf{J}_u}{k-(1+\alpha)d}+(1-\mathbf{J}_u) \right)
\end{displaymath}
\end{lemma}
The proof is a bit standard, e.g., see \cite{old-GnpSampling}. For the sake of completeness we provide one in Section \ref{sec:lemma:DisProbBound}.

So as to bound the expected number of disagreements in the coupling above we use disagreement percolation.
Consider the configuration space  ${\cal D}=\{ {\tt agree }, {\tt disagree} \}^{V(B)}$ and let
${\rho }:{\cal D}\to [0,1]$ be the product probability measure on $\cal D$ such that 
 each $u\in B$ is set ``disagree" with probability $\rho_u$.

Let  $Z\in {\cal D}$ be distributed as in $\rho$. A path of disagreement in $Z$ is a path in $B$ such that 
for every vertex $u$ in the path it holds $Z(u)={ \tt disagree}$.
Let $R_i=R_i(Z)$ be the number of paths  of disagreements of   length $i$ that 
start from the  vertex $r$. It holds that

\begin{equation}\label{eq:prop:ExpctNoDisBlock:Basis}
\mathbb{E}[ H(X_{t+1}(B), Y_{t+1}(B) ) | X_t, Y_t, B_t=B  ]  \leq \sum_{i\geq 0}\mathbb{E} [R_i].
\end{equation}
The proposition will follow by bounding appropriately  $\mathbb{E}[R_i]$.
For this,  we need to consider the tree of self-avoiding walks $T=T(B_t)$;
$T$ is rooted at vertex $r$. At its level $j$, $T$ contains a copy of every  vertex in $B$ which is reachable
from $r$ with a path of length $j$.

For any two vertices $u, v\in T$, let ${\cal P}(v, u)$ be  the probability that the  path that connect $u$ and $v$  is a path of 
disagreement in $Z$. From the linearity of expectation, for   i=0,1,2\ldots, we have that
\begin{equation} \label{eq:ExpctRiBasis}
\mathbb{E}[R_i]=\sum_{u \ : \ {\tt dist}(r,u)=i} {\cal P}(r, u)   
\end{equation}
Note that the sum runs over the vertices at distance $i$ from $v$,  the root of $T$.
We have that
\begin{eqnarray}
{\cal P}(r, v) & = & \prod_{u\in L(r,v)}\rho_u=
\prod_{u\in L(r, v)}\left( \frac{2\mathbf{J}_{u} }{k-(1+\alpha)d}+
(1-\mathbf{J}_u)\right) \nonumber \\
&\leq & \prod_{u\in L(r, v)}\left( \frac{2\mathbf{J}_{u} }{11d/2-(1+\alpha)d}+
(1-\mathbf{J}_u)\right),  \nonumber
\end{eqnarray}
where $L(r, v)$ is the path that connects $v$ to $r$. The last inequality uses the
fact that $k\geq 11d/2$ while the product is decreasing with $k$.
Pugging the above into \eqref{eq:ExpctRiBasis} we get that
\[
\mathbb{E}[R_i] \leq \sum_{w \ : \ {\tt dist}(r,w)=i \ \ }  \prod_{u\in L(r,w)}\left( \frac{2\mathbf{J}_{u} }{11d/2-(1+\alpha)d}+
(1-\mathbf{J}_u)\right).
\]

\noindent
We use the  the following lemma  to bound $\mathbb{E}[R_i]$.

\begin{lemma}\label{lemma:TGrowthRate}
Let $\alpha,\gamma, c$ and $d$ be as in the statement of Proposition \ref{prop:ExpctNoDisBlock}. Then, it holds that
\begin{equation}\label{eq:BoundedGrowth}
\mathbb{E}[R_i] \leq \frac{(1-\theta)^i}{11d/2-(1+\alpha)d} \qquad i=0,1,\ldots,
\end{equation}
where 
 $\theta\leq \min \left \{1-\frac{(1+\gamma)(1+\alpha)}{11/2-(1+\alpha)}, 1- \left( \frac{(1+\alpha)}{11/2-(1+\alpha)}\right)^{9/10} \right\}$.
\end{lemma}
The proof of Lemma \ref{lemma:TGrowthRate} appears in Section \ref{sec:lemma:TGrowthRate}.

\remove{

At this point, observe the following: Let $P_v$ be a path in $T$ from the root $v_0$ to some vertex $v$. 
The probability that $P_v$ is a path of disagreement is upper bounded by the quantity ${\cal C}_{p,s}(P_v)$
where $p=\frac{2}{\frac{11}{2}d-(1+\alpha)d}=\frac{1}{2.245d}$, $s=(1+\alpha)d=1.01d$. Thus, it holds that
$$
E_{\cal P}[Z_i]\leq \sum_{v\in l_i(T)}{\cal C}_{p,s}(P_v),
$$
where $l_i(T)$ is the set of vertices at level $i$ in $T$. 
Also, for the tree $T$ a condition similar to (\ref{eq:GrowthCondition}) holds, if we set $\delta=d$ and $\zeta=\gamma$.
I.e. each $P_v$ in $T$ satisfies the condition (\ref{eq:GrowthCondition}) as the disagreeing vertex $w$ is a break-point 
and the root of $T$ is adjacent to $w$. Also, it holds that $sp,\delta p\in [\frac{1}{100(1+\zeta)}, \frac{1}{1+\zeta}]$.
Lemma \ref{lemma:TGrowthRate}, then,  implies that $E_{\cal P}[Z_i]\leq p(1-\theta)^i$. 
}

So as to bound $\mathbb{E} [R_i]$  we take the maximum  possible value for $\theta$, w.r.t. the parameters $d, \alpha, c, \gamma$. It is direct  that $\theta\leq 0.5127$, 
that is 
\[
\mathbb{E}[R_i]\leq \left( {0.44543}/{d}\right)\ \left(0.4873\right)^{i}.
\]
The proposition follows by plugging the above into \eqref{eq:prop:ExpctNoDisBlock:Basis}.

\subsection{Proof of Lemma \ref{lemma:DisProbBound}}\label{sec:lemma:DisProbBound}

For the moment assume that the block $B_t$ is a tree. 
Assume, also, that the vertex $u$ is the {\em first vertex} that is going 
to be coloured in the coupling. That is, $u$ is next to the disagreeing vertex $w\in \partial B_t$.
Let 
\[
S_X=\{\sigma\in [k]^{V(B_t)} \ | \  \Pr[X_{t+1}(B_t)=\sigma|X_{t}, Y_t ]>0 \ \textrm{and}\ \sigma(u)\neq Y_t(w) \}.
\]
$S_X$ contains all the legal  colourings which can be assigned  to $X_{t+1}(B_t)$
such that  $u$ takes a  colouring which is different than $Y_t(w)$.
Similarly, we have
\[
S_Y=\{\sigma\in [k]^{V(B_t)} \ | \  \Pr[Y_{t+1}(B_t)=\sigma|X_{t}, Y_t ]>0 \ \textrm{and}\ \sigma(u)\neq X_t(w) \}.
\]  
It is trivial to verify that  $S_Y$ and $S_X$ are {\em identical}. This implies that
there is a coupling such that 
\[
\Pr[X_{t+1}(u) = Y_{t+1}(u)| X_{t+1}(u), Y_{t+1}(u)\notin \{X_{t}(w),Y_{t}(w)\}]=1.
\]
Then we get that
\begin{eqnarray}
\Pr[X_{t+1}(u) \neq  Y_{t+1}(u)]&=&\Pr[X_{t+1}(u), Y_{t+1}(u)\in \{X_t(w),Y_t(w)\}]\nonumber \\
&\leq &\max\left\{\Pr[X_{t+1}(u)=Y_{t}(w)], \Pr[Y_{t+1}(u)=X_t(w)] \right\}.\nonumber
\end{eqnarray}
The last inequality follows from a maximal coupling of $X_{t+1}(u), Y_{t+1}(u)$. 

Noting that for $u$ it holds ${\tt degree}(u)<k$,  it is not hard to see that in the previous inequality the  maximum  can be greater than
$\frac{1}{k-{\tt degree}(u)}$. If $k <{\tt degree}(u)$,  the maximum can be  bounded by 1.

If $u$ is not the first vertex to be coloured in the coupling, the situation is essentially the same. Let ${\cal C}(u)$ 
be the set of  already coloured vertices just before colouring $u$. Observe that if it is impossible to find a vertex 
which is adjacent to a  disagreement then for the rest vertices we have identical coupling. 
Assume that ${\cal C}(u)$ contains  disagreeing vertices.  
Since we always pick first a vertex next to a disagreement, ${\cal C}(u)$ must be a connected subtree of $B_t$.

Consider the subtree in $B_t\setminus {\cal C}(u)$ which contains vertex $u$. 
Let us call this subtree $T_u$.
Then the previous arguments apply directly to $T_u$ and we get the same bounds for 
$\Pr[X_{t+1}(u) \neq  Y_{t+1}(u)]$.

For the case where $B$ is a unicyclic graph, the only difference is  that there are at most {\em two paths}
 from which the disagreement can reach the vertex $u$ (as opposed to one 
path in the case of trees).  This implies that, in the worst case, $X_{t+1}(u)$ and $Y_{t+1}(u)$ should avoid 
at most two colours so as there is no disagreement. I.e. for appropriate $c_1,c_2,c_3,c_4\in [k]$,
it holds that
\[
\Pr[X_{t+1}(u) \neq  Y_{t+1}(u)] \leq \max\left\{ \Pr[X_{t+1}(u)\in \{c_1,c_2\}], \Pr[Y_{t+1}(u)\in {c_3,c_4}] \right\}.
\]
If ${\tt degree}(u)<k-1$, then the r.h.s. of the inequality above is upper bounded by $\frac{2}{k-{\tt degree}(u)}$.
Otherwise, i.e. ${\tt degree}(u)\geq k-1$, we set the trivial bound $1$.

\subsection{Proof of Lemma \ref{lemma:TGrowthRate}}\label{sec:lemma:TGrowthRate}

\begin{lemmaproof}{\ref{lemma:TGrowthRate}}
For the sake of brevity let $q=2/(11d/2-(1+\alpha)d)$ and $f=(1+\alpha)d$.
Let us introduce some further notation.
Given a vertex $u$ in $T$, let $T_u$ be the subtree of $T$ defined as follows:
In $T$ delete the edge that connect $u$ to its parent. Then, $T_u$ is the
tree that contains vertex $u$. We always imply that the root of $T_u$ is the
vertex $u$.
Finally,  for some integer $j$ and some tree $T'$, we let ${\cal E}(T', j)$ be the set of
vertices at level $j$ in $T'$.

Since the vertex $r$ is right next to a disagreement, the construction of blocks ensures that ${\tt degree}(r)\leq f$. 
Then ${\cal P}(r,r)=q$. i.e.   (\ref{eq:BoundedGrowth}) is indeed true for $i=0$.

Consider  $i>0$.    It holds that
\begin{equation}
\sum_{v \ : \ {\cal E}(T, i)}
{\cal P}(r, v) =
\rho(r) \times \sum_{u \ : \ \textrm{child of $r$}\ } \left(\sum_{v\in {\cal E}(T_{u}, i-1)  } {\cal P}(u, v)  \right).
\label{eq:SubTreeReduction}
\end{equation}
Using  \eqref{eq:SubTreeReduction} we can get  a new set of  conditions which imply  (\ref{eq:BoundedGrowth}). 
That is, for each child of the root $u$  it should hold that
\begin{equation}
\sum_{v\in {\cal E}(T_u, {i-1}) } {\cal P}(u, v)  \leq q(1-\theta)^i\times \frac{1}{\rho(r)\ {\tt degree}(r)}. \label{eq:SubTreeCondition}
\end{equation}
If the bound in (\ref{eq:SubTreeCondition}) holds for every subtree rooted at a child of the root, 
then  we can plug it  into (\ref{eq:SubTreeReduction}) and get (\ref{eq:BoundedGrowth}).
We can repeat  the same argument for every subtree $T_u$ and then to the subsequent subtrees and
so on. 
In particular, after having repeated the above argument $i$ times, we deal  subtrees  $T_u$, where $u\in {\cal E}(T, i)$.
For each such $T_u$ it turns out that we  need to  bound ${\cal P}(u,u)$.
That is, the condition we have is the following: for every $u \in {\cal E}(T, i)$ it should hold that
\begin{eqnarray}
{\cal P}(u,u) &\leq& q(1-\theta)^i\times \prod_{u'\in L(r,u) } \left( \rho(u')\ {\tt degree}(u') \right)^{-1}\nonumber\\
&=&  q(1-\theta)^i\times \left(\frac{1}{qf}\right)^{i-t}\times \prod_{v \in M\backslash\{u\}}\frac{1}{{\tt degree}(v)},
\label{eq:RLvDef}
\end{eqnarray}
where the set $M$ contains all the vertices in $L(r,u)$ of degree larger than $f$, while $t=| M\backslash\{u\}|$.

To this end, observe that we have that ${\cal P}(u,u)=\rho_u$.
Clearly,  this observation  reduces the problem of showing \eqref{eq:RLvDef}  to 
\begin{equation}\label{eq:UnfoldedInduction}
q(1-\theta)^i\times \left(\frac{1}{qf}\right)^{i-t} \times \left( \frac{\mathbf{J}_{u }}{q}+1-\mathbf{J}_{u }\right)\times 
\prod_{v \in M\backslash\{u\}}\frac{1}{{\tt degree}(v)}
 \geq 1,
\end{equation}
for every $u\in {\cal E}(T,i)$.

Note that if there is $\theta$ such that  \eqref{eq:UnfoldedInduction} is true for every $u\in {\cal E}(T, i)$, then our arguments imply that 
\eqref{eq:BoundedGrowth} is also true. The lemma follows by showing that such $\theta$ exists.

So as to deal with \eqref{eq:UnfoldedInduction} it crucial  to upper bound 
$\prod_{v \in M\backslash\{u\}} {{\tt degree}(v)}$. At this point we need to use the weighting schema. In particular,
we use the fact that the vertex $r$, the root of $T$, is next to a break-point. This assumption implies the
following: For every $u\in {\cal E}(T, i)$
\begin{equation}\label{eq:WeightConditionForRi}
\prod_{u' \in L(r, v)} \left ( \frac{\mathbf{J}_{u'}}{1+\gamma}  +d^{c}(1-\mathbf{J}_{u'})  \right) \leq (1+\gamma),
\end{equation}
In the following series of claims, we exploit \eqref{eq:WeightConditionForRi} to show that indeed
there is $\theta$ that satisfies \eqref{eq:UnfoldedInduction}.

\begin{claim}\label{claim:CaseA-TGrowthRate}
If ${\tt degree}(u)\leq f$, then   the condition in (\ref{eq:UnfoldedInduction}) is true for 
$\theta\leq 1-qf(1+\gamma)$.
\end{claim}

\begin{claim}\label{claim:CaseB-TGrowthRate}
If ${\tt degree}(u)> f$ and $|M\backslash\{u\}|=0$, then the condition in (\ref{eq:UnfoldedInduction}) is true for 
$\theta\leq 1-(qf)^{9/10}$.
\end{claim}

\begin{claim}\label{claim:CaseC-TGrowthRate}
If ${\tt degree}(u)> f$ and $|M\backslash\{u\}|\geq 1$, then  the condition in (\ref{eq:UnfoldedInduction}) is true  for 
$\theta\leq 1-qf(1+\gamma)$.
\end{claim}
Taking  $\theta\leq \min\{1-qf(1+\gamma),1-(qf)^{9/10}\}$, the above claims imply that \eqref{eq:UnfoldedInduction}
is true.  The lemma follows.
\end{lemmaproof}
\vspace{.3cm}

\begin{claimproof}{\ref{claim:CaseA-TGrowthRate}}
For each path $L(r, u)$, where $u\in {\cal E}(T, i)$ the condition \eqref{eq:WeightConditionForRi} implies that 
\begin{equation}
\frac{1}{(1+\gamma)^{i-t}}\prod_{v \in M\setminus \{u\}}
\left( d^{10} \  {\tt degree}(v) \right) \leq (1+\gamma). \label{CaseAWeightStatus}
\end{equation}
For the sake of brevity let the l.h.s. of \eqref{eq:UnfoldedInduction} be denoted as ${\cal Q}(u)$.

If $t\geq 1$, we have that
\begin{eqnarray}
{\cal Q}(u) &=& q(1-\theta)^i\frac{1}{(qf)^{i-t}}\frac{1}{q} \prod_{v \in M\setminus\{u\} } \left({\tt degree}(v)\right)^{-1}
\nonumber \\
&\geq& (1-\theta)^i\left(\frac{1}{qf(1+\gamma)}\right)^{i-t}\ \frac{d^{10t}}{1+\gamma}
\qquad \qquad \mbox{[we used \eqref{CaseAWeightStatus}]}
\nonumber \\
&\geq& \left(\frac{1-\theta}{qf(1+\gamma)}\right)^{i}\frac{\left ( d^{10}qf(1+\gamma)\right)^t}{1+\gamma} 
  \nonumber\\
&\geq & \left(\frac{1-\theta}{qf(1+\gamma)}\right)^{i},
\nonumber 
\end{eqnarray}
where the last inequality follows from the fact for large $d$ we have that $\frac{\left [d^{10}qf(1+\gamma)\right]^t}{1+\gamma}\gg 1$.

If $t=0$, then it is direct to show that 
\[
{\cal Q}(u)=q(1-\theta)^i\frac{1}{(qf)^{i-t}}\frac{1}{q} \prod_{v \in M\setminus\{u\} } \left({\tt degree}(v)\right)^{-1}=\left(\frac{1-\theta}{qf}\right)^i.
\]
For both cases the claim is true.
\end{claimproof}
\vspace{.3cm}

\begin{claimproof}{\ref{claim:CaseB-TGrowthRate}}
Since  condition  \eqref{eq:WeightConditionForRi} holds for $T$, our assumptions
implies that  $\frac{d^{10}}{(1+\gamma)^{i}}  \cdot {\tt degree}(u) \leq 1+\gamma$.
Then, we get a lower bound on $i$,  i.e. it holds that $i\geq \frac{\ln ( d^{10}{\tt degree}(u)) }{\ln(1+\gamma)}-1$.

For the sake of brevity, we denote the l.h.s. of \eqref{eq:UnfoldedInduction} be denoted as ${\cal Q}(u)$.
We have that
\[
{\cal Q}(u)
= q(1-\theta)^i\frac{1}{(qf)^{i}}\geq \left( \frac{q^{10}}{(qf)^{i}} \right)^{1/10}\cdot\left(\frac{1-\theta}{(qf)^{9/10}}\right)^i.
\]
The claim will follow by showing that $(qf)^{i}<q^{10}$. 

In what follows  we use the facts 
$qf\geq 1/e $,  $\gamma<10^{-2}$ and  $i\geq \frac{\ln[d^{10} {\tt degree}(u)]}{\ln(1+\gamma)}-1$.
That is, 
\begin{eqnarray}
(qf)^{i}&\leq &(qf)^{ \frac{\ln (d^{10} {\tt degree}(u))}{\ln (1+\gamma)}-1 }
\ \leq \ (d^{10} {\tt degree}(u))^{- \frac{\ln (1/qf)}{\ln\left({1+\gamma}\right)}} \  (qf)^{-1} 
\nonumber \\
&\leq& (d^{10} {\tt degree}(u))^{-1}\ (qf)^{-1}  \nonumber\\
&\leq& \left( {d^{10} \sqrt{ {\tt degree}(u)} }\right)^{-1}\  \left((qf)\  \sqrt{ {\tt degree}(u)}\right)^{-1} \leq q^{10}.\nonumber
\end{eqnarray}
The last inequality holds for large $d$.
The claim follows.
\end{claimproof}
\vspace{.3cm}

\begin{claimproof}{\ref{claim:CaseC-TGrowthRate}}
With the assumption of the claim,  the condition in (\ref{eq:WeightConditionForRi}) 
implies that 
\begin{equation}\label{eq:CaseCWeightStatus}
\frac{1}{(1+\gamma)^{i-t}}\prod_{v \in M\backslash\{u\}}d^{10} \cdot {\tt degree}(v) \leq 1+\gamma . 
\end{equation}
where $t\geq 1$, i.e. $M\backslash\{u\}\neq \emptyset$. 

In the following derivations we use the, easy to verify,  fact that
$\frac{q}{1+\gamma}[d^{10} qf(1+\gamma)]^t\gg 1$.
For the sake of brevity, we denote the l.h.s. of \eqref{eq:UnfoldedInduction} by ${\cal Q}(u)$.
We have that
\begin{eqnarray}
{\cal Q}(u)&=&q(1-\theta)^i\frac{1}{(qf)^{i-t}} \prod_{v \in M\backslash\{u\}} \left( {\tt degree}(v)\right)^{-1}\nonumber \\
&\geq &
{q}\left( {1+\gamma}\right)^{-1}\  (1-\theta)^i\left({qf(1+\gamma)}\right)^{-(i-t)}d^{10t} 
\hspace{4.75cm} \mbox{[we use \eqref{eq:CaseCWeightStatus}]}
\nonumber \\
&\geq& {q}\left( {1+\gamma}\right)^{-1}\ (1-\theta)^i\left( {qf(1+\gamma)}\right)^{-i} [d^{10}qf(1+\gamma)]^t
\nonumber \\
&\geq& \left(\frac{1-\theta}{qf(1+\gamma)}\right)^{i}.
\hspace{4.38cm} \left[\textrm{since } \frac{q}{1+\gamma}(d^{10}qf(1+\gamma))^t\gg 1\right]
\nonumber 
\end{eqnarray}
The claim follows.
\end{claimproof}

\section{Proof of Theorem \ref{thrm:IS-ExctDisagreement}}\label{sec:thrm:IS-ExctDisagreement}

The coupling is such that we choose the same block to update in both $(X_t)_{t\geq 0}$ and 
$(Y_t)_{t\geq 0}$. Let $B_t$ be the block whose colouring is chosen to be updated in the 
coupling at time $t$.

Assume that $X_t\oplus Y_t=\{w\}$, i.e. the two copies disagree only on vertex $w$.
Let $B_w$ be the block that contains  the disagreeing vertex $w$. Also,  let $\Gamma_w \subset {\cal B}\setminus \{B_w\}$ be 
the blocks that are adjacent to $w$.

To simplify our further analysis we need to make the following observations:
\begin{description}
\item [Observation 1.] If $B_t=B_w$, then there is a coupling such that $H(X_{t+1}, Y_{t+1})=0$.
\item [Observation 2.] If $B_t \notin \Gamma_w\cup B_w$, then there is a coupling such that $H(X_{t+1}, Y_{t+1})=1$.
\item [Observation 3.] If $B_t=\{u\}\in \Gamma_w$, then $u$ is a break-point and $\Pr[X_{t+1}(B_t)\neq Y_{t+1}(B_t)]\leq \frac{\lambda}{\lambda+1}$.
\end{description}
We get  Observation 1 and Observation 2 with exactly the same arguments as in the proof of Theorem \ref{thrm:ExctDisagreement}, 
i.e.  the assumption that $X_t\oplus Y_{t}=\{w\}$  implies that  $X_t(\partial B_t)=Y_t(\partial B_t)$, 
whenever $B_t\in {\cal B}\setminus \Gamma_w$.

As far as Observation 3 is regarded, w.l.o.g. assume that $X_{t}(w)$ is {\em occupied},
i.e. $w$ belongs to the independent set,  and $Y_{t}(w)$ is {\em unoccupied}.
Clearly $X_{t+1}(u)$ cannot become occupied. The only way we can have disagreement at
$u$, is when all the neighbours of $u$, apart from $w$, in both configurations  are unoccupied. 
Then, $Y_{t+1}(u)$ becomes occupied (disagreeing)  with probability $\frac{\lambda}{1+\lambda}$.

The following proposition, somehow, generalizes Observation 3, in the sense that it considers
general blocks from $\Gamma_w$ and not necessarily single vertex ones. Note that $\Gamma_w$
contains multi-vertex blocks when $w$ is a  single vertex block.

\begin{proposition}\label{prop:IS-ExpctNoDisBlock}
Let $\epsilon,\alpha,\gamma,c$ and $\lambda$ be as in the statement of Theorem \ref{thrm:IS-ExctDisagreement}. 
There is a coupling such that
\[
\mathbb{E}[ H(X_{t+1}(B_t), Y_{t+1}(B_t) )\  |\  X_t, Y_t, B_t\in \Gamma_w ] \leq \left(1-(5/4)\epsilon_0\right) / {d}. 
\]
\end{proposition}
The proof of Proposition \ref{prop:IS-ExpctNoDisBlock} appears in Section \ref{sec:prop:IS-ExpctNoDisBlock}.

In what follows,  we let $N=|{\cal B}|$, i.e. each $B\in {\cal B}$ is chosen with probability  $1/N$.
So as to prove \eqref{eq:ContrCond} we  consider a number of different cases for $B_w$ and the position
of $w$ inside $B_w$, similarly to those in the proof of Theorem \ref{thrm:ExctDisagreement}
in Section \ref{sec:thrm:ExctDisagreement}.\\ \vspace{-.3cm}

\noindent
{\tt Case 1:}
The  block $B_w$ is multi-vertex and $w$   is not adjacent to any vertex in $\partial B_w$.
Since $w$ is inside $B_w$ we have that $\Gamma_w=\emptyset$.
There is a coupling which has the following properties:
With probability  $1/N$  we have $B_t=B_w$.  Then, using Observation 1, we get that $H(X_{t+1},Y_{t+1})=0$.
Also, with  probability  $1-1/N$ we have $B_t\notin \Gamma_w\cup B_w$, since $\Gamma_w=\emptyset$. 
Then, from Observation 2,  we get that  $H(X_{t+1},Y_{t+1})=1$
Combining the above,  we get that 
\[
\mathbb{E}  [H(X_{t+1},Y_{t+1}) \ |\  X_t, Y_t ]=1-{1}/{N}. 
\]

\noindent
{\tt Case 2:}
The block $B_w$  is multi-vertex   and   $w$  is  adjacent to at least one vertex in $\partial B_w$.

Recalling the properties of $B\in \cal B$, our assumptions 
imply the following:
we have that ${\tt degree}(w)\leq (1+\alpha)d$.
Also,  $\Gamma_w\neq \emptyset$, while each  $B\in \Gamma_w$  is single vertex ones.
i.e. $B$ is the block of a  break-point.

There is a coupling which has the following properties:
With probability $1/N$ we have $B_t=B_w$ and, according to Observation 1, we have
$H(X_{t+1}, Y_{t+1})=0$.
Also, with probability $|\Gamma_w|/N$, we have that $B_t\in \Gamma_w$. 
Since each $B\in \Gamma_w$ consists of a single break-point, 
Observation 3 implies that $\Pr[X_{t+1}(B_t)\neq Y_{t+1}(B+t)]\leq \frac{\lambda}{1+\lambda }$.
Finally, if $B_t \notin \Gamma_w\cup B_w$, then,  Observation 2,
implies that $H(X_{t+1}, Y_{t+1})=1$. 
With the above arguments, we get that
\[
\mathbb{E} [H(X_{t+1},Y_{t+1})\  |\  X_t, Y_t]\leq 1-{1}/(2N), 
\]
since $\alpha\leq \epsilon_0/3$ and $\lambda \leq (1-\epsilon)/(2d)$.

\noindent
{\tt Case 3:}
$B_w$ is a single vertex block.   i.e. $w$ is a break-point.

With probability $1/N$ we have $B_t=B_w$  which implies that  $H(X_{t+1},Y_{t+1})=0$.
Also, with probability  $|\Gamma_w| /N$ we have that $B_t\in \Gamma_w$.
Then using Proposition \ref{prop:IS-ExpctNoDisBlock} we have that the number of
disagreements generated is  at most $(1-\frac54\epsilon_0)/d$.
Finally, if  $B_t\notin \Gamma_w \cup B_w$, we have that   $H(X_{t+1},Y_{t+1})$=1.
Combining all the above together, we get that 
\[
\mathbb{E}  [H(X_{t+1},Y_{t+1}) \ | \  X_t, Y_t] \ \leq \ 1-\frac{1}{N}+\frac{|\Gamma_w|}{N} \left (1- \frac54\epsilon_0\right)\frac{1}{d}
\ \leq \ 1-\frac{\epsilon_0}{2N},  
\]
where the above holds since $|\Gamma_w|\leq (1+\alpha)d$ and $\alpha\leq \epsilon_0/3$.
The theorem follows.

\subsection{Proof of Proposition \ref{prop:IS-ExpctNoDisBlock}}\label{sec:prop:IS-ExpctNoDisBlock}

The proof of this proposition is very similar to the proof of Proposition
\ref{prop:ExpctNoDisBlock} for the colouring model. 
Consider some $B\in \Gamma_w$. Recall that $X_t(w)\neq Y_t(w)$  and this is the only disagreement at
the boundary of $B$. Also, there is exactly one vertex in $B$ which is adjacent to $w$. Let us call this vertex
$r$. For each vertex $u\in B$ let $\mathbf{J}_u$ be an indicator which is 1 if ${\tt degree}(u)\leq (1+\alpha)d$ 
and $0$, otherwise.

We wish to couple $X_{t+1}(B)$ and $Y_{t+1}(B)$ as close as possible.
Identical coupling is  precluded for the whole block $B$ due to the disagreement at the boundary of $B$, recall that $w\in \partial B$.
The coupling we use for setting $X_{t+1}(B)$ and $Y_{t+1}(B)$ decides the colour assignment of each
vertex $u\in B$ at a time.  Every time we give priority to the vertices which are next to a disagreement,
i.e. we decide the colouring for these vertices first.  If there are vertices in $B$ whose colouring is not decided
but non of them is next to a disagreeing vertex, then it is easy to see that we can colour them using identical
coupling.

\begin{corollary}\label{lemma:DisProbBoundHC}
Assume in the above process of deciding $X_{t+1}(B)$ and $Y_{t+1}(B)$, the process consider some
vertex $u\in B$.  There is a coupling  such that $u$ becomes disagreeing with probability  at most 
$ \rho_u=\frac{\lambda}{1+\lambda}$
\end{corollary}

\noindent
So as to bound the expected number of disagreements in the coupling above we use disagreement percolation
from \cite{DisPerc}.  Consider the configuration space  ${\cal D}=\{ {\tt agree }, {\tt disagree} \}^{V(B)}$ and let
${\rho }:{\cal D}\to [0,1]$ be the product probability measure on $\cal D$ such that 
 each $u\in B$ is set ``disagree" with probability $\rho_u$.

Consider a configuration space  ${\cal D}=\{ {\tt  agree}, {\tt disagree} \}^{V(B_t)}$ and let
${\rho }:{\cal D}\to [0,1]$ be the product probability measure on $\cal D$ such that 
 each $u\in B_t$ is set ``disagree" with probability $\rho_u=\frac{\lambda}{1+\lambda}$.
 We get this value for $\rho_u$ by working  as in the case of a  single vertex update case of 
 Theorem \ref{thrm:IS-ExctDisagreement}.

Let  $Z\in {\cal D}$ be distributed as in $\rho$. A path of disagreement in $Z$ is a path in $B$ such that 
for every vertex $u$ in the path it holds $Z(u)={ \tt disagree}$.
Let $R_i=R_i(Z)$ be the number of paths  of disagreements of   length $i$ that 
start from the  vertex $r$. It holds that

\begin{equation}\label{eq:prop:IS-ExpctNoDisBlock:Basis}
\mathbb{E}[ H(X_{t+1}(B_t), Y_{t+1}(B_t) ) \ |\  X_t, Y_t, B_t\in \Gamma_w ]  \leq \sum_{i\geq 0}\mathbb{E} [R_i],
\end{equation}

\noindent
For bounding $\mathbb{E}[R_i]$ we need to consider the tree of self-avoiding walks $T=T(B_t)$: 
$T$ is rooted at vertex $r$. At level $j$, $T$ contains a copy of a vertex in $B$ which is reachable
from $r$ with a path of length $j$

For any two vertices $u, v\in T$, let ${\cal P}(v, u)$ be  the probability that the 
path that connect $u$ and $v$  is a path of disagreement in $Z$.
From the linearity of expectation, for   i=0,1,2\ldots, we have that
\begin{equation} \label{eq:IS-ExpctRiBasis}
\mathbb{E}[R_i]=\sum_{u \ : \ {\tt dist}(r,u)=i} {\cal P}(r, u)   
\end{equation}
Note that the sum runs over the vertices at distance $i$ from $v$,  the root of $T$.

We have that
\[
{\cal P}(r, v)=\prod_{u\in L(r,v)}\rho_u=
\prod_{u\in L(r, v)} \frac{\lambda}{1+\lambda},
\]
where $L(r, v)$ is the path that connects $v$ to $r$,  the root of $T$.

Pugging the above into \eqref{eq:IS-ExpctRiBasis} we get that
\[
\mathbb{E}[R_i]=\sum_{w \ : \ {\tt dist}(r,w)=i \ \ }  \prod_{u\in L(r, w)} \frac{\lambda}{1+\lambda}
\ \leq \ \sum_{w \ : \ {\tt dist}(r,w)=i \ \ }  \prod_{u\in L(r, w)} \frac{\lambda_0}{1+\lambda_0},
\]
where $\lambda_0=(1-\epsilon_0)/(2d)$ and $\epsilon_0=\min\{\epsilon, 10^{-2}\}$. 
The last derivation follows from the observation that $\lambda_0\geq \lambda$ and the fact that
$f(x)=\frac{x}{1+x}$ is increasing.

We use the  the following lemma  to bound $\mathbb{E}[R_i]$.

\begin{lemma} \label{lemma:IS-TGrowthRate}
Let $\epsilon, \alpha,\gamma, c$ and $d$ be as in the statement of Proposition \ref{prop:IS-ExpctNoDisBlock}. 
Also, let $\epsilon_0=\min\{\epsilon, 10^{-2}\}$ and let $\lambda_0=(1-\epsilon_0)/(2d)$.
It holds that
\begin{equation}\label{eq:IS-BoundedGrowth}
\mathbb{E}[R_i] \leq  \frac{\lambda_0}{1+\lambda_0} (1-\theta)^i  \qquad i=0,1,\ldots
\end{equation}
where  for any  $\theta \leq 1-\frac{ \lambda_0}{1+\lambda_0}(1+\alpha)(1+\gamma)d$.
\end{lemma}
The proof of Lemma \ref{lemma:IS-TGrowthRate} appears in Section \ref{sec:lemma:IS-TGrowthRate}.

\remove{
\begin{lemma}
Consider a tree $H=(V,E)$ containing vertices of different degrees.
Let $l_i$ denote the set of vertices at the level $i$ in $H$. For a
vertex $v\in H$ let $L_v$ denote the simple path connecting it
to the root $r$. 
For $L_v$ we define the following weight:
$$
{\cal C}_{p,s}(L_v)=p^{|L_v|},
$$
where $p\in [0,1]$. Given $d,s>0$, for any vertex 
$v$ such that $\Delta(v)>s$ the following condition holds: 
\begin{equation}\label{eq:IS-GrowthCondition}
\prod_{u\in L_v}\left(\frac{\mathbb{I}_{\{\Delta(u)\leq s\}}}{(1+\gamma)}+d^{10}\cdot \Delta(u)\cdot\mathbb{I}_{\{\Delta(u)>s\}}\right)\leq (1+\gamma),
\end{equation}
for some $\gamma>0$. 
Assume that $d,s$ are sufficiently large and $d,s\gg (1+\gamma)$,
while both $(s\cdot p)$, $(d\cdot p)\in [\frac{1}{100}\frac{1}{1+\gamma},\frac{1}{1+\gamma}]$.
Then (\ref{eq:IS-GrowthCondition}) implies that 
\begin{equation}\label{eq:IS-BoundedGrowth}
\sum_{v\in l_i}{\cal C}_{p,s}(L_v)\leq p \cdot \left(1-\theta \right)^{i} \qquad \forall i\geq 0,
\end{equation}
for any $\theta \leq 1-ps(1+\gamma)$.
\end{lemma}
The proof of Lemma \ref{lemma:IS-TGrowthRate} appears in Section \ref{sec:lemma:IS-TGrowthRate}.
}

Using  Lemma \ref{lemma:IS-TGrowthRate} and setting 
$\theta= 1-\frac{\lambda_0}{1+\lambda_0}(1+\gamma)(1+\alpha)d \geq \frac{1+\epsilon_0/2}{2}$.
We get that 
\[
\sum_{i\geq 0}\mathbb{E}[R_i] \leq  \theta^{-1} {\lambda_0}/({1+\lambda_0})
 \leq \left[ 1-(5/4)\epsilon_0\right] /d.
\]
The proposition follows by plugging the above into \eqref{eq:prop:IS-ExpctNoDisBlock:Basis}.

\subsection{Proof of Lemma \ref{lemma:IS-TGrowthRate}}\label{sec:lemma:IS-TGrowthRate}

\begin{lemmaproof}{\ref{lemma:IS-TGrowthRate}}
For the sake of brevity we let $q=\frac{\lambda_0}{1+\lambda_0}$ and $f=(1+\alpha)d$.
Let us introduce some further notation, similar to Lemma \ref{lemma:TGrowthRate}. Given a vertex $u$ in 
$T$, let $T_u$ be the subtree of $T$ defined as follows: In $T$ delete the edge that connect $u$ with its parent. 
Then, $T_u$ is the  tree that contains vertex $u$. We always imply that the root of $T_u$ is the
vertex $u$.  Finally,  for some integer $j$ and some tree $T'$, we let ${\cal E}(T', j)$ be the set of
vertices at level $j$ in $T'$.

Since the vertex $r$ is right next to a disagreement, the construction of blocks ensures that ${\tt degree}(r)\leq f$. 
Then ${\cal P}(r,r)=q$. i.e.   (\ref{eq:IS-BoundedGrowth}) is indeed true for $i=0$.

Consider now that $i>0$.  It holds that
\begin{equation}
\sum_{v \ : \ {\cal E}(T, i)}
{\cal P}(r, v) =
\rho(r) \times \sum_{u \ : \ \textrm{child of $r$}\ } \left(\sum_{v\in {\cal E}(T_{u}, i-1)  } {\cal P}(u, v)  \right).
\label{eq:IS-SubTreeReduction}
\end{equation}
Using  \eqref{eq:IS-SubTreeReduction} we can get a new set of  conditions
which imply  (\ref{eq:IS-BoundedGrowth}). That is,
for each child of the root $u$  it should hold that
\begin{equation}
\sum_{v\in {\cal E}(T_u, {i-1}) } {\cal P}(u, v)  \leq q(1-\theta)^i\times \frac{1}{\rho(r)\ {\tt degree}(r)}. \label{eq:IS-SubTreeCondition}
\end{equation}
If the bound in \eqref{eq:IS-SubTreeCondition} holds for every subtree rooted at a child of the root, 
then  we can plug it  into \eqref{eq:IS-SubTreeReduction} and get \eqref{eq:IS-BoundedGrowth}.
We can repeat  the same argument for every subtree $T_u$ and then to the subsequent subtrees and
so on. 
After having repeat the above argument $i$ times, we deal  subtrees  $T_u$, where $u\in {\cal E}(T, i)$.
For each such $T_u$ it turns out that we  need to  bound ${\cal P}(u,u)$.
That is, the condition we have is as follows: for every $u \in {\cal E}(T, i)$ it should hold that
\begin{eqnarray}
{\cal P}(u,u) &\leq& q(1-\theta)^i \prod_{v\in L(r, u)} \left(\rho(v) \ {\tt degree}(v) \right)^{-1}\nonumber \\
&\leq & q(1-\theta)^i \times \left(\frac{1}{qf}\right)^{i-t}\times \prod_{v \in M\backslash\{u\}}\frac{1}{{\tt degree}(v)},
\label{eq:IS-RLvDef}
\end{eqnarray}
where the set $M$ contains all the vertices in $L(r,u)$ of degree larger than $f$, while $t$ is the
cardinality of $M\backslash\{u\}$.
Observe that we have that ${\cal P}(u,u)=q=\frac{\lambda_0}{1+\lambda_0}$.
This   reduces the problem of showing \eqref{eq:IS-RLvDef}  to 
\begin{equation}\label{eq:IS-UnfoldedInduction}
q(1-\theta)^i\times \left(\frac{1}{qf}\right)^{i-t}\times \prod_{v \in M\backslash\{u\}}\frac{1}{{\tt degree}(v)}
\times \frac{1}{q}  \geq 1,
\end{equation}
for every $u\in {\cal E}(T,i)$.

Note that if we find $\theta$ such that \eqref{eq:IS-UnfoldedInduction} is true for every $u\in {\cal E}(T, i)$, then our 
arguments imply that  \eqref{eq:IS-BoundedGrowth} is also true. The lemma follows by showing that 
such $\theta$ exists.

A main technical challenge is to get an appropriate bound for the product of degrees of the vertices in
$M\setminus\{u\}$ in \eqref{eq:IS-UnfoldedInduction}. For this endeavor a useful observation is that $r$ is next to $w$, which
is a break-point. This implies that for every vertex $v$ in $T$ we have that
\begin{equation}\label{eq:IS-WeightConditionForRi}
\prod_{u\in L(r, v)} \left ( \frac{\mathbf{J}_u}{1+\gamma}  +d^{c}(1-\mathbf{J}_u)  \right) \leq (1+\gamma),
\end{equation}
where $L(r, v)$ is the path that connects $v$ to $r$, the root of $T$.

We consider two cases, the first one is $t\geq 1$ and the second one is $t=0$.
For the sake of brevity we denote the l.h.s. of \eqref{eq:IS-UnfoldedInduction} by
 ${\cal Q}(u)$.

Assume that  $t\geq 1$. Then,  \eqref{eq:IS-WeightConditionForRi} implies that
\begin{equation}\label{eq:local-IS-WeightConditionForRi}
\frac{1}{(1+\gamma)^{i-t}}\prod_{v \in M\setminus\{u\}} \left( d^{10} \  {\tt degree}(v) \right) \leq (1+\gamma). 
\end{equation}
Using this fact we have that
\begin{eqnarray}
{\cal Q}(u)&=&q(1-\theta)^i\frac{1}{(qf)^{i-t}}\frac{1}{q} \prod_{v\in M\backslash\{u\}} \frac{1}{{\tt degree}(v)} \nonumber \\
&\geq &(1-\theta)^i\left(\frac{1}{qf(1+\gamma)}\right)^{i-t}\frac{d^{10t}}{1+\gamma} 
\hspace{4cm} \mbox{[we use \eqref{eq:local-IS-WeightConditionForRi}]}
\nonumber \\
&\geq& \left(\frac{1-\theta}{qf(1+\gamma)}\right)^{i}\frac{\left [d^{10}qf(1+\gamma)\right]^t}{1+\gamma}
\geq \left(\frac{1-\theta}{qf(1+\gamma)}\right)^{i},
\nonumber 
\end{eqnarray}
where the last inequality follows from the fact that $\frac{\left [d^{10}qf(1+\gamma)\right]^t}{1+\gamma}\gg 1$
for large $d$.

If $q=0$, then it is direct to see that ${\cal Q}(u)=\left(\frac{1-\theta}{qf}\right)^i$.
Clearly,  \eqref{eq:IS-UnfoldedInduction} is true by taking  $\theta\leq 1-qf(1+\gamma)$. 
The lemma follows.
\end{lemmaproof}

\section{Proof of Theorem \ref{thrm:BreakablePaths}}\label{sec:thrm:BreakablePaths}

Consider the graph $\G$ and the weighting schema defined in Section \ref{sec:technique}. Recall that 
each vertex $u$  is assigned weight $W(u)$ such that
 
\begin{equation}\nonumber
 W(u)=\left \{
\begin{array}{lcl}
(1+\gamma)^{-1} &\qquad& \textrm{if ${\tt degree}(u)\leq (1+\alpha)d$}\\ \vspace{-.3cm} \\ 
d^c  \  {\tt degree}(u) && \textrm{otherwise},
\end{array}
\right .
\end{equation}
where  $\alpha,\gamma,c>0$ are specified in the statement of Theorem \ref{thrm:BreakablePaths}.

Let $\varrho_{\ell}$ be the probability that for an elementary path
of legth $\ell$, in $\G$, to  have a no  break point.  
Noting that the number of elementary paths of length $\ell $ in $\G$ is at most the number of paths 
(not necessarily elementary) of length $\ell$, we get that  
\begin{eqnarray}
\Pr[\mathbf{U} \neq \emptyset]&\leq& \mathbb{E}[|\mathbf{U}|] 
\leq {n \choose \ell+1}\  \left(\frac{d}{n}\right)^{\ell }\  \varrho_{\ell}
\nonumber
\hspace{1.35cm} \mbox{[by the linearity of expectation]}\\
&\leq& \left(\frac{ne}{\ell+1}\right)^{\ell+1}\  \left(\frac{d}{n}\right)^{\ell}\  \varrho_{\ell}\nonumber
\hspace{2.2cm}  \left[ \textrm{as ${n\choose i}\leq (ne/i)^i$}\right] 
\\ 
&\leq& nd^{\ell}\ \varrho_{\ell} \leq n^{2} \  \varrho_{\ell}
\hspace{3.4cm}
\left[\textrm{ as $\ell={\ln n}/{ (\ln d)^{5}}$}\right].
\label{eq:ObjectiveWithVarrho}
\end{eqnarray}
The theorem follows by bounding appropriately  $\varrho_{\ell}$.   Before proceeding, perhaps, it is useful to clarify what
$\varrho_{\ell}$ is  in  more technical terms. Consider some permutation of $\ell$ vertices $L=w_{1},\ldots, w_{\ell}$. 
Let  ${\cal I}_{L}$ be the event that  $L$ is an elementary path in $\G$.  Then, it holds that
\begin{displaymath}
\varrho_{\ell}=\Pr[\textrm{there are no break points in $L$}\  |\  {\cal I}_{L}].
\end{displaymath}

\begin{definition}
For an elementary path $L=w_1, \ldots, w_{\ell}$, a vertex $w_i\in L$ is called {\em left-break} or {\em right-break} for $L$
if it has the corresponding property below: \\ \vspace{-.1cm}\\
\begin{tabular}{@{}r@{}@{}r@{}c@{}}
 \hspace*{0.4cm}& {\bf left-break: } &
There is no path $L'\in {\cal P}(w_i)$ such that $\prod_{u \in L'}W(u)>1$ 
and $L'\cap L$ contains only $v_j$ s.t.  $j\leq i$, \\ \vspace{-.2cm}\\
 \hspace*{0.4cm}& {\bf right-break: } &
There is no path $L'\in {\cal P}(w_i)$ such that $\prod_{u \in L'}W(u)>1$ 
and $L'\cap L$ contains only $v_j$ s.t.  $j\geq i$. \\
\end{tabular}
\end{definition}
Recall  that for a  vertex $v$,  ${\cal P}(v)$ denotes the set of all paths of length at most $\frac{\ln n}{d^{2/5}}$ that start from 
$v$.
\begin{remark}\label{remark:RightLeftVsBreakPoing}
Each  $w_i \in L$ which is both left and right break  is also a break point.
\end{remark}

\begin{proposition}\label{prop:NoOfSpecialR}
Let $d, \alpha, \gamma, c$ and  $ \ell$ be as in the statement of Theorem \ref{thrm:BreakablePaths}. 
Let ${\cal S}$ be the set of elementary paths in $\G$ of length $\ell$ such that for each $P\in {\cal S}$
both the number of left-breaks and right breaks is at least $0.9\ell$.
Let  $L=w_1, \ldots, w_{\ell}$ be a permutation of vertices of $G$.
It holds that 
\[
\Pr[ L\notin {\cal S}\ | \  {\cal I}_L ]\leq    4n^{-\frac{1}{2} \ln d },
\]
where  ${\cal I}_{L}$ is  the event that   $L$  is an elementary path in $\G$.
\end{proposition}
The proof of Proposition \ref{prop:NoOfSpecialR} appears in 
Section \ref{sec:prop:NoOfSpecialR}.

Note that every path in ${\cal S}$ has at least $0.8\ell$ vertices which are both left-break and right-break. 
Then, Remark \ref{remark:RightLeftVsBreakPoing} implies  that each path in $ {\cal S}$ 
has at least $0.8\ell$ break points. With this observation we have that 
\[
 \varrho_{\ell} \leq \Pr[L\notin {\cal S }\  | \ {\cal I}_L ] \leq  4n^{-\frac{1}{2} \ln d }, 
\]
where $L=(w_1,\ldots, w_{\ell})$ is a permutation of $\ell$ vertices

The theorem follows by plugging the above inequality into (\ref{eq:ObjectiveWithVarrho}).

\subsection{Proof of Proposition \ref{prop:NoOfSpecialR}}\label{sec:prop:NoOfSpecialR}

Let ${\cal S}_r$ be the set of length $\ell$ elementary paths in $\G$ which have at least $0.9\ell$ right-breaks.
Similarly,  ${\cal S}_l$ be the set of length $\ell$ elementary paths in $\G$ which have at least $0.9\ell$ left-breaks.
Note that ${\cal S}={\cal S}_r\cap {\cal S}_l$.

It is direct  that the events $L\in {\cal S}_l$ and $L\in {\cal S}_r$  are symmetric.
Using this observation we get that
\begin{eqnarray}
\Pr[L\notin {\cal S} \ |\  {\cal I}_L ] &\leq & \Pr[L\notin S_r \cup  L\notin S_l \ |\  {\cal I}_L  ] \nonumber \\
&\leq & \Pr[L \notin S_r  \:|\:{\cal I}_L ]+\Pr[L\notin S_l  \:|\: {\cal I}_L]  \qquad \mbox{[union bound]}\nonumber \\
&\leq & 2 \Pr[L\notin {\cal S}_l \:|\: {\cal I}_L ]. \hspace{3.325cm} \mbox{[due to symmetry]}
\end{eqnarray}

\noindent
The proposition will   follow by bounding appropriately  $\Pr[L\notin {\cal S}_l \:|\: {\cal I}_L ]$, 
i.e.  we will show that:
\begin{equation}
\Pr[L\notin {\cal S}_l \:|\: {\cal I}_L ] \leq  2n^{-\frac{1}{2} \ln d } . \label{eq:YlTailBound}
\end{equation}

\noindent
Consider the weighting schema for the vertices of $\G$ we defined in Section \ref{sec:thrm:BreakablePaths}.
Using this weighting of the vertices and $L$ we define a new schema as follows:
For each $w_i\in L$ let $N_i$ denote the set of the vertices outside $L$  which are adjacent to $v_i$. For every $u \in N_i$ 
let $E_{out}(u)$ denote the influence on vertex $w$, only from paths of length  at most $\ln n/d^{2/5}$ that do not use 
vertices in $L$. 
For every $w_i$ let 
\[
Q(w_i)=\max_{u \in N_i}\left\{E_{out}(u )\right\}. 	
\]
Now, we associate each vertex in $L$ with the (new) weight
\begin{displaymath}
 U(w_i)=\left \{
\begin{array}{lcl}
\frac{\max\left\{1, Q(w_i)\right\}}{1+\gamma}&\qquad& \textrm{if ${\tt degree}(w_i)\leq (1+\alpha)d$}\\ \vspace{-.1cm}
\max\left\{1, Q(w_i)\right\}\  d^c \  {\tt degree}(w_i) && \textrm{otherwise,}
\end{array}
\right .
\end{displaymath}
where $\alpha, \gamma, c$ are defined in the statement of the proposition.

By definition  any vertex $w_i$ such that $U(w_i)>1$ cannot be a left break for $L$. Let 
\begin{equation}\label{eq:HSetDef}
C=\{ i \in [\ell] \; |\; U(w_i)>1\}.
\end{equation}
For each $j\in C$, let ${R}_j=w_{j},\ldots,w_{s}$  be the {\em maximal} subpath of $L$ such that 
for any $j'\in [j,s]$  it holds $\prod_{r=j}^{j'}U(w_{r})>1$.

\begin{lemma}\label{lemma:RvsRBreaks}
In $L$,  a vertex $w_{i}\notin C$ is a left  break if there is no $w_j\in C$ such that $j\leq i$ and $w_{i}\in {R}_{j}$.
\end{lemma}
The proof of Lemma \ref{lemma:RvsRBreaks} appears in Section \ref{sec:lemma:RvsRBreaks}.

Let ${ R}=\cup_{j \in C} { R}_j$.    Lemma \ref{lemma:RvsRBreaks} implies that  the number of left-breaks of $L$ is equal to  
$\ell -|{ R}|$.  Using this observation, it suffices to show the following proposition whose proof appears in 
Section \ref{sec:prop:tRMgfBound}.

\begin{proposition}\label{prop:tRMgfBound}
Let  $d,  \alpha,\gamma,c, \ell$  and $L$ be as in  Proposition \ref{prop:NoOfSpecialR}.  
It holds that
\[
\Pr\left[|{ R}|\geq 0.1\ell   \:|\: {\cal I}_L\right] \leq  2n^{-\frac{1}{2} \ln d }.
\]
\end{proposition}

\noindent
The proposition follows.

\section{Proof of Proposition \ref{prop:tRMgfBound}}\label{sec:prop:tRMgfBound}

Before proceeding with the proof of Proposition \ref{prop:tRMgfBound} we need to introduce few concepts.
\begin{definition}
Let  $L=w_1, \ldots, w_{\ell}$ be a permutation of vertices. for every $i=1,\ldots, \ell$,  let   
$\psi_{i}:\mathbb{R}_{\geq 0}\to [0,1]$ be the marginal distribution of  $U(w_i)$  induced by the graph $
\G$.
Also, let $\xi:\mathbb{R}^{\ell}_{\geq 0}\to[0,1]$ be the joint distribution of
$(U(w_1), \ldots, U(w_{\ell})$.
Finally, let  $\zeta:\mathbb{R}^{\ell}_{\geq 0}\to [0,1]$ be such that $\zeta=\otimes^{\ell}_{i=1}\psi_i$, 
where $\otimes$ stands for the tensor product.
\end{definition}

\noindent
That is, for $\mathbf{v}$ distributed as in $\zeta$, the components of the vector are independent with each other.
Note that for the aforementioned distributions we do necessarily  have that  $w_1,\ldots,w_{\ell}$ is a path.

For every $\mathbf{a}\in \mathbb{R}^{\ell}_{\geq 0}$  let 
\[
{\cal C}(\mathbf{a})=\{ i\in [\ell]\; |\; \mathbf{a}(i)>1\}.
\]
Furthermore, for each $j\in {\cal C}$, let ${\cal Q}_j(\mathbf{a})=\mathbf{a}(j), \mathbf{a}(j+1),\ldots, \mathbf{a}(s)$  be  
a {\em maximal} sequence  such that  for any $j'\in [j,s]$  it holds $\prod_{r=j}^{j'} \mathbf{a}(r)>1$.
Also,   let ${\cal Q}(\mathbf{a})=\bigcup_{j}{\cal Q}_j(\mathbf{a})$.

\begin{remark}\label{remark:RVsQ}
Note that for the random variable $R$ we have that ${R}={\cal Q}(\mathbf{u})$, where $\mathbf{u}$ 
is distributed as in $\xi$.
\end{remark}

\noindent
A function ${\cal H}:\mathbb{R}^{\ell}_{\geq 0}\to \mathbb{R}_{\geq 0}$
is called    {\em increasing} if for any $\mathbf{a,b}\in\mathbb{R}^{\ell}_{\geq 0}$ such 
that  $\mathbf{a}(i)\geq \mathbf{b}(i)$,   $ \forall i\in [\ell]$, it holds that ${\cal H}(\mathbf{a})\geq {\cal H}(\mathbf{b})$.

\begin{lemma}\label{lemma:MonotonicityReductionIndependence}
Let  $d, \alpha,\gamma,c, \ell$ and $L$ be  as in the statement of Proposition \ref{prop:tRMgfBound}.
Let $\mathbf{u}, \mathbf{v}$ be distributed as in $\xi$ and $\zeta$, respectively. 
Then,   for any  increasing function ${\cal H}:\mathbb{R}^{\ell}_{\geq 0}\to \mathbb{R}_{\geq 0}$,
we have that
\[
\mathbb{E} \left [{\cal H}( \mathbf{u}) \ | \  {\cal I}_L  \right ] \leq 2 \ 
\mathbb{E} \left [{\cal H}( \mathbf{v}) \ | \  {\cal F}_L  \right ],
\]
where   ${\cal F}_{L}$ (resp.  ${\cal I}_{L}$) indicates  the event that  in $\G$, the graph that induces $\mathbf{u,v}$,
$L$ is a path (resp.  elementary path).
\end{lemma}

\begin{remark}
In the expression $\mathbb{E}  [{\cal H}(\mathbf{v}) \ | \  {\cal F}_L   ] $, conditioning on
${\cal F}_L$ does not introduce any correlation between $\mathbf{v}(1), \mathbf{v}(2), \ldots \mathbf{v}(\ell)$. 
It only specifies that the distribution of  $\mathbf{v}(i)$ is induced by $\G$ conditional that $L$ is a path. 
\end{remark}

\noindent
The proof of Lemma \ref{lemma:MonotonicityReductionIndependence} appears in Section \ref{sec:lemma:ProdFactWeightB}.

Let the indicator $\mathbf{1}\{|{R}| \geq 0.1\ell\}$ be equal to 1 if the  the event $|{R}|\geq 0.1\ell$
holds and zero otherwise.
It holds that $\mathbf{1}\{|{R}| \geq 0.1\ell\}$ is an increasing function of $U(w_1), \ldots, U(w_{\ell})$.
Then,  Lemma \ref{lemma:MonotonicityReductionIndependence} and Remark \ref{remark:RVsQ}
imply the following corollary.

\begin{corollary}
Let $\mathbf{v}$ be distributed as in $\zeta$. It holds that
\[
\Pr\left[|{ R}|\geq 0.1\ell   \:|\: {\cal I}_L\right] 
 \leq 2\  \Pr\left[ |{\cal Q}(\mathbf{v})|\geq 0.1\ell   \:|\: {\cal F}_L\right].
\]
\end{corollary} 

\remove{
\begin{corollary}
Let $\mathbf{v}$ be distributed as in $\zeta$. For any $t\geq 0$ we have that
\[
 \mathbb{E}\left[e^{t|{R}|}\:|\:  {\cal I}_{L}  \right]\leq 2\  \mathbb{E}\left[e^{t |{\cal Q}(\mathbf{v})|}\:|\:  {\cal E}_{L}  \right].
\]
\end{corollary}
}

\noindent
The proposition will follow by bounding appropriately $\Pr\left[ |{\cal Q}(\mathbf{v})|\geq 0.1\ell   \:|\: {\cal F}_L\right]$.

Given  $\mathbf{a}\in \mathbb{R}^{\ell}_{\geq 0}$, consider the set ${\cal C}(\mathbf{a})$,
as defined above.  For any  $j<s$ such that ${\cal Q}_{j}(\mathbf{a}) \cap {\cal Q}_{s}(\mathbf{a}) \neq \emptyset$ 
both  ${\cal Q}_{j},  {\cal Q}_{s}$   have the same endpoint, e.g.,    if  ${\cal Q}_{s}=\mathbf{a}(s),\ldots, \mathbf{a}(i)$, 
then  ${\cal Q}_{j}=\mathbf{a}(j), \ldots, \mathbf{a}(i)$.
Let 
$$
{\cal J}(\mathbf{a})=\left\{j\in {\cal C}(\mathbf{a}) \; | \; 
\nexists s<j: {\cal Q}_{j}\subset {\cal Q}_{s}\right\}.
$$
For any $j,s\in {\cal J}(\mathbf{a})$ it holds that ${\cal Q}_j(\mathbf{a})\cap {\cal Q}_{s}(\mathbf{a})=\emptyset$, 
while  $|{\cal Q}(\mathbf{a})|=\sum_{j\in {\cal J}(\mathbf{a})}| {\cal Q}_{j}(\mathbf{a})|$.

Let  $\mathbf{v}$ be distributed as in $\zeta$. Let  ${\cal M}$ be the event that 
$|{\cal J}(\mathbf{v})|\geq (d^{-7/10}\ln d) \ln n$. For any $t>0$, it holds that
\begin{eqnarray}
\Pr[|{\cal Q}(\mathbf{v})|\geq 0.1\ell \ |\  {\cal F}_L] &=& \Pr\left [ \exp(t |{\cal Q}(\mathbf{v})|) \geq \exp(t\ell/10)\ | \ {\cal F}_L\right] \nonumber \\
&\leq &\Pr[{\cal M}^c\ | \  {\cal F}_L] + \Pr\left [ \exp(t |{\cal Q}(\mathbf{v})|) \geq \exp(t\ell/10)\ | \ {\cal F}_L, {\cal M}\right] 
\nonumber \\
&\leq &\Pr[{\cal M}^c\ | \  {\cal F}_L] + 
\exp(-t\ell/10) \ \mathbb{E}\left [ \exp(t |{\cal Q}(\mathbf{v})|)  \ | \ {\cal F}_L, {\cal M}\right] . \label{eq:Target4Prop:tRMgfBound}
\end{eqnarray}

\noindent
We need to bound appropriately  the quantities on the r.h.s. of \eqref{eq:Target4Prop:tRMgfBound}.
For this, we use of the following tail bound for the distribution $\zeta$.

\begin{theorem} \label{thrm:AcCL-Tail}
Let  $\alpha\in (0,3/2)$,  $0< \gamma< 4, c>0$  and  $\delta>0$ be fixed numbers.
Consider  $\G=G(n,d/n)$ for sufficiently large $d$. Let $L=w_1,\ldots, w_{l }$ be permutation of 
vertices of $\G$ such that  $l \leq  \frac{\ln n}{(\ln d)^2}$. 

Let $\mathbf{v}\in \mathbb{R}^{l }_{\geq 0}$ be distributed as in $\zeta$. 
Then, for any $r, q\in [\ell]$ s.t. $r+q-1\leq l $ we have that
\[
\Pr\left [  \prod^{r+q-1}_{j=q} \mathbf{v}(j) \geq \delta \;|\;  {\cal F}_L \right ]\leq 2\exp\left[-d^{7/10}\left(r +\ln \delta\right) \right].
\] 
\end{theorem}
The proof of Theorem \ref{thrm:AcCL-Tail} appears in Section \ref{sec:thrm:AcCL-Tail}.

As far as bounding $\Pr[{\cal M}^c\ | \ {\cal F}_L]$ is regarded we use the following result.

\begin{lemma}\label{lemma:UpperBoundOnJ}
Let  $\alpha, \gamma,c, d,  \ell $ and $L$ be as in the statement of Proposition \ref{prop:tRMgfBound}.
Let $\mathbf{v}$ be distributed as in $\zeta$.
 It holds that 
\[
\Pr  [|{\cal J}(\mathbf{v})| \geq ({d^{-7/10}}{\ln d} ) \ln n \ | \ {\cal F}_L] \leq  n^{-\frac{1}{2}\ln d}.
\]
\end{lemma}
The proof of Lemma \ref{lemma:UpperBoundOnJ} appears in Section \ref{sec:lemma:UpperBoundOnJ}.

We use the following sequence of results to show that $\mathbf{E}[\exp(t|{\cal Q}(\mathbf{v})|)\ | \ {\cal F}_L, {\cal M}]$
is sufficiently small. 
When there is no danger of confusion,  we abbreviate   $\mathbb{E}[\cdot|  {\cal F}_L ]$ and
$\Pr[\cdot | {\cal F}_L ]$ to  $\mathbb{E}_{ {\cal F}}$ and  $\textrm{P}_{ {\cal F} }[\cdot]$, respectively.
Similarly, we abbreviate ${\cal J}(\mathbf{v}), {\cal Q}(\mathbf{v})$ and ${\cal Q}_j(\mathbf{v})$ to 
${\cal J}, {\cal Q}$ and ${\cal Q}_j$, respectively.

\begin{lemma}\label{lemma:exptRj-bounding}
Let  $\alpha, \gamma,c, d$ and $\ell$ be as in the statement of Proposition \ref{prop:tRMgfBound}.
Let $\mathbf{v}$ be distributed as in $\zeta$. For any $j\in {\cal C}(\mathbf{v})$,
it holds that
\[
 \mathbb{E}_{\cal F} \left[\exp( t\ | {\cal Q}_{j}(\mathbf{v})|)  \right]\leq  2\exp(d).
\]
\end{lemma}
The proof of Lemma \ref{lemma:exptRj-bounding} appears in Section \ref{sec:lemma:exptRj-bounding}.

\begin{claim}\label{claim:NegativeAssociation}
Let $\mathbf{v}$ be distributed as in $\zeta$.  For any integer $s\geq 0$ and for any $j\in {\cal C}(\mathbf{v})$ 
it holds that
\[
\mathbb{E}_{\cal F}\left[ \exp(t|{\cal Q}_{j}|) \  | \ |{\cal Q}_j|\leq s \right] \leq \mathbb{E}_{\cal F}\left[ \exp(t|{\cal Q}_{j}|) \right].
\]
\end{claim}
The above claim is elementary to prove. For this reason we omit the proof.

For $j=1, \ldots, |\cal J|$,  let $i_j$ be the $j$-th smallest element in ${\cal J}$, e.g., $i_1$ is the smallest element,
$i_2$ is the second smallest element and so on.  Using Lemma \ref{lemma:exptRj-bounding} and Claim \ref{claim:NegativeAssociation}, 
we have that

\begin{eqnarray}
\mathbb{E}_{ {\cal F}} \left [\exp\left(t | {\cal Q}|\right)  \ | \ {\cal J}\right]
 &=&  \mathbb{E}_{ {\cal F}} \left [\exp\left(t | {\cal Q}|\right) \  |\  i_j \ \textrm{for } j=1,2, \ldots, |{\cal J}|  \right]  \nonumber \\ 
 &=& 
\prod^{|J|}_{j=1} \mathbb{E}_{ {\cal F}} \left [\exp\left(t | {\cal Q}_{i_j} |\right) \  |\  \  i_j \ \textrm{for } j=1,2, \ldots, |{\cal J}| \right] 
\label{eq:FactorisationForQjs} \\
 &=& 
\prod^{|J|}_{j=1} \mathbb{E}_{ {\cal F}} \left [\exp\left(t | {\cal Q}_{i_j} |\right) \  |\   | {\cal Q}_{i_j} | < | i_{j+1}-j_{j} | -1\right] 
\nonumber \\
 &\leq & 
\prod^{|J|}_{j=1} \mathbb{E}_{ {\cal F}} \left [\exp\left(t | {\cal Q}_{i_j} |\right)  \right]  
\hspace{3cm}\mbox{[from Claim \ref{claim:NegativeAssociation}]}
\nonumber \\
 &\leq & [2\exp(d)]^{|{\cal J}|}.  \hspace{4.1cm} \mbox{[from Lemma \ref{lemma:exptRj-bounding}]}
 \label{eq:ExpTQCondJ}
\end{eqnarray}

\noindent
Note that \eqref{eq:FactorisationForQjs} holds because of the independence of $\mathbf{v}(i)$s. In particular, 
conditional on $i_{j}$, for $j=1, \ldots, |{\cal J}|$, for any two $s, t$ such that $s\leq i_j\leq t$  it holds that
$\mathbf{v}(s)$ and $\mathbf{v}(t)$ are independent with each other.
From  \eqref{eq:ExpTQCondJ} we get that
\begin{eqnarray}
\mathbb{E}\left[ \exp(t|{\cal Q}(\mathbf{v})|) \ | \ {\cal F}_L, {\cal M} \right]  & \leq & 
 \mathbb{E}\left[  \left(2e^{d} \right)^{|\cal J|} | \ {\cal F}_L, {\cal M} \right]  \ \leq \ 
    n^{2d^{3/10}\ln d}, \label{eq:EexpTQBound}
\end{eqnarray}
where in the last derivation of we use the fact that, conditional on $\cal M$, we have 
$|{\cal J}|\leq d^{-7/10}\ln d \ln n$.

Plugging \eqref{eq:EexpTQBound} and the bound from Lemma \ref{lemma:UpperBoundOnJ} into 
\eqref{eq:Target4Prop:tRMgfBound}  and then setting $t=\sqrt{d}$ we get the desirable bound for 
$\Pr[{\cal Q}(\mathbf{v}) \geq 0.1\ell \ | \ {\cal F}_L]$.  The proposition follows.

\subsection{Proof of Lemma \ref{lemma:MonotonicityReductionIndependence}}\label{sec:lemma:ProdFactWeightB}

\newcommand\N{\mathbold{N}}

\begin{lemmaproof}{ \ref{lemma:MonotonicityReductionIndependence}}
In what follows,  let  ${\cal E}_L$ be the event that $L$ is path in $\G$ (not necessarily elementary).

Consider $\G$ conditional on event ${\cal E}_L$. Let $\N$  be the subgraph  of $\G$ defined as follows:
For each  $w_i\in L$ we  define  the sets ${Z}_{i,s} \subseteq V(\G)$,  for $s=0,\ldots, h$, 
where $h=10\ln n/d^{2/5} $.  Let  ${ Z}_{i,0}=\{w_i\}$. For $s>1$, we get    ${ Z}_{i,s}$ by working inductively. 
Let ${\Gamma}_{i,s} \subset V(\G)$ contain all the vertices  but those  which belong to  $L$
and those which belong to $\bigcup_{j <  i}\bigcup_{j'\leq h}{ Z}_{j,j'} $ and $\bigcup_{j'<s}{ Z}_{i,j'}$. 
Then $Z_{i,s}$ contains all the vertices in ${\Gamma}_{i,s}$ which have  a neighbour in $Z_{i,s-1}$.
For $i=1,\ldots, \ell$,  let   ${N}_{i,h}$ be the induced subgraph of $\G$ with vertex set   $\bigcup^h_{s=0}{ Z}_{i,s}$.  
Then the subgraph $\N$ is the union  $(\cup^{\ell}_{i=1}{N}_{i,h}) \cup L$.

The construction of $\N$ guarantees that apart from the edges of $L$,   there is no other  edge with ends
in   $N_{i,h}$ and $N_{j,h}$, for $i\neq j$.
That is, even if there are such edges in $\G$ these are neither  revealed during the construction of $\N$ nor
included in $\N$. It is elementary  that in  $\N$ there is no (extra) path $P'$ of length less than $10 (d^{-9/10})\ln n $ 
that connects any two vertices in $L$. That is, in $\N$ the path $L$ is elementary.

Let  ${\cal X}$ be the set of  edges of $\G$ that connect any two $N_{i,h}$ and $N_{j,h}$, for $j\neq i$.
Note that our construction guarantees that if ${\cal X}$ is empty then $L$ is elementary.
The contrary does not necessarily holds, since some edges in $\cal X$ introduce a path
between two vertices of $L$ which is of length greater than $10\ln n/d^{9/10}$.

Let ${\cal Y}\subseteq {\cal X}$ contain precisely each edge such that, its appearance
in   $\G$ makes $L$ non elementary.  That is if ${\cal Y}= \emptyset$ if and only if
$L$ is elementary. We  argue  that, typically, ${\cal Y}$is empty. In particular we have the following result,
whose proof appears after this proof.

\begin{claim}\label{claim:EmptySetOfCrossingsTail}
It holds that $\Pr[  {\cal Y}=\emptyset \ | \  {\cal E}_L ]\geq 3/4$.
\end{claim}

\noindent
Consider the second weighting schema, we introduced in Section \ref{prop:NoOfSpecialR}, w.r.t. the path $L$ in the subgraph $\N$. Let $\bar{U}(w_i)$ be the
weight assigned to each vertex $w_i\in L$. 
Since $\N$ is a subgraph of $\G$, then for every $w_i\in L$ it holds that
\begin{equation}\label{eq:MonotonicityOfSchema2}
\bar{U}(w_i)\leq U(w_i).
\end{equation}
We note   that if ${\cal Y}=\emptyset$, then in \eqref{eq:MonotonicityOfSchema2} we have  equality.
That is, for  $\mathbf{u}=[U(w_1),\ldots, U(w_{\ell}) ]$ and   ${\mathbf{w}}=[\bar{U}(w_1),\ldots, \bar{U}(w_{\ell}) ]$
we have that 
\begin{equation}\label{eq:Reduction2SubgraphPathWeight2}
\mathbb{E} \left [   {\cal H}( \mathbf{u} ) \  | \  {\cal I}_L  \right] =
\mathbb{E}[  {\cal H}( {\mathbf{w}}  )  \   | \   {\cal E}_L, {\cal Y}=\emptyset].
\end{equation}

\noindent
Furthermore,  we have that
\begin{eqnarray}
\mathbb{E}[ {\cal H}( {\mathbf{w}}  )  \   | \  {\cal E}_L, {\cal Y}=\emptyset]
\leq 
( {\Pr[ { \cal Y}=\emptyset \ | \   {\cal E}_L] } )^{-1}\  \mathbb{E}[  {\cal H}( {\mathbf{w}}  ) \  | \   {\cal E}_L]
\; \leq \; (4/3) \ \mathbb{E}[ {\cal H}( {\mathbf{w}}  ) \   |\   {\cal E}_L ], \nonumber 
\end{eqnarray}
where in the last inequality we used  Claim \ref{claim:EmptySetOfCrossingsTail}. That is, 
\begin{equation}\label{eq:UnConditionCrossingSetAndMAxDegree}
\mathbb{E} \left [   {\cal H}( \mathbf{u} ) \ |\  {\cal I}_L  \right]  \leq 
(4/3) \ \mathbb{E}[ {\cal H}( {\mathbf{w}}  ) \  | \ {\cal E}_L], 
\end{equation}

\noindent
Let $\mathbf{v}\in \mathbb{R}^{\ell}$ be distributed as in $\zeta$. We are going to show that
\begin{equation}\label{eq:SubgraphVsIndependentWeights}
\mathbb{E}[{\cal H}(\mathbf{w}) \ | \ {\cal E}_L] \leq \mathbb{E}[{\cal H}(\mathbf{v}) \ |\  {\cal F}_L].
\end{equation}
Then, the lemma will follow by combining \eqref{eq:SubgraphVsIndependentWeights} and \eqref{eq:UnConditionCrossingSetAndMAxDegree}.

Let $\mathbf{w}^j=[ \mathbf{v}(1), \ldots,  \mathbf{v}(j),  \mathbf{w}(j+1)) , \ldots, \mathbf{w}(\ell)]$, 
for any $j=0,\ldots, \ell$. 
Then,  \eqref{eq:SubgraphVsIndependentWeights} follows by  showing that 
 \begin{eqnarray}\label{eq:StepWiseStochasticOrder}
\mathbb{E}\left [ {\cal H}( \mathbf{w}^{j}  ) \  |\  {\cal F}_L, {\cal I}_l \right] \leq 
\mathbb{E} \left [ {\cal H}( \mathbf{w}^{j+1}  ) \   | \  {\cal F}_L, {\cal I}_L \right]  
\qquad \forall j=0,\ldots, \ell-1,
\end{eqnarray}
where the conditioning on the  events  ${\cal F}_L, {\cal I}_L $, imply that the instance of $\G$ which specify
the vectors  $\mathbf{w}$ and $\mathbf{v}$, have path $L$.
Note that  $\mathbf{w}^0=\mathbf{w}$ and $\mathbf{w}^{\ell}=\mathbf{v}$.

Since ${\cal H}$ is increasing,  for \eqref{eq:StepWiseStochasticOrder} it suffices to show that, for every
$j=0,\ldots, \ell-1$, there is a  coupling for $\mathbf{w}^{j}$ and $\mathbf{w}^{j+1}$ such that $\mathbf{w}^{j}(i)\geq \mathbf{w}^{j+1}(i)$, 
for every $i=1,\ldots, \ell$.

There is a coupling such that the first $j$ components of $\mathbf{w}^{j}$ and $\mathbf{w}^{j+1}$ are identical.
This follows from the fact that the $\mathbf{w}^{j}(i)$ and $\mathbf{w}^{j+1}(i)$ are identically distributed and
independent of the other components, for $i=1,\ldots, j$.
Due to  \eqref{eq:MonotonicityOfSchema2}, we have that $\mathbf{v}(j+1)$ dominates $\mathbf{w}(j+1)$.  This implies that
there is a coupling such that $\mathbf{w}^{j}(j+1) \leq \mathbf{w}^{j+1}(j+1)$.
Finally,   there is a coupling such that $\mathbf{w}^{j}(i) =  \mathbf{w}^{j+1}(i)$, for $i>j+1$.
The above arguments completes the proof of \eqref{eq:StepWiseStochasticOrder}. 
The lemma follows.
\end{lemmaproof}
\vspace{.3cm}

\begin{claimproof}{\ref{claim:EmptySetOfCrossingsTail}}
Each   $u\in { Z}_{i,s-1}$ has a  number of neighbours in  ${\Gamma}_{i,s}$ whose distribution is
dominated by ${\cal B}(n,d/n)$. This implies that
\[
\mathbb{E}[|V(\N)|  \ | \ {\cal E}_L] \leq  \frac{d^{h+1}-1}{d-1}\ell \  \leq \ n^{d^{-0.3}}. 
\qquad \mbox{[as $h=10\ln n/d^{2/5}$ and $\ell=\Theta(\ln n)$]}
\]

\noindent
Let $B$ be the event that $ |V(\N)|\geq 1000n^{d^{-0.3}}$.  
 Markov's inequlity implies that
\begin{equation}\label{eq:EmptySetOfCrossingsTailBound4B}
\Pr[B \  |\  {\cal E}_P]\leq 10^{-3}.
\end{equation}
Recall that ${\cal X}$ is the set of edges in $\G$ which have  one end in $N_{i,h}$ and the other end
at $N_{j,h}$, for every $i,j$ such that $i\neq j$. Then, we have that
\begin{eqnarray}
\Pr[{\cal Y} \neq \emptyset \ | \ {\cal E}_L] &\leq & \Pr[ |{\cal X}| \geq 1 \  |\  {\cal E}_L]  \nonumber \\
&\leq & \Pr[B \  | \ {\cal E}_L]+\Pr[  |{\cal X}| \geq 1 \  |\   B^c, {\cal E}_L]\nonumber \\
&\leq & 10^{-3}+\mathbb{E}[|{\cal X}| \  |\   B^c, {\cal E}_L], \label{eq:EmptySetOfCrossings1821} 
\end{eqnarray}
where in the last inequality we use \eqref{eq:EmptySetOfCrossingsTailBound4B} for the first probability term
and Markov's inequality for the second one.

Each potential edge between $N_{i,h}$ and $N_{j,h}$ appears with probability $d/n$. 
Conditioning on $B^c$ implies that there are at most $10^6n^{2d^{-0.3 }}$
potential edges. Then, we get that 
$$
\mathbb{E}[ |{\cal X}|  \  |\   B^c, {\cal E}_L] = O(n^{-1/2}).$$
The claim follows by plugging the bound for $\mathbb{E}[|{\cal X}| \ | \ B^c, {\cal E}_L]$ into \eqref{eq:EmptySetOfCrossings1821}.
\end{claimproof}

\subsection{Proof of Lemma \ref{lemma:UpperBoundOnJ}}\label{sec:lemma:UpperBoundOnJ}

Let $N=d^{-7/10} {\ln d}  \ln n $. Since,  it holds that $|{\cal J}(\mathbf{v}) |\leq | {\cal C}(\mathbf{v}) |$,
the lemma will follow by showing that
\[
\Pr\left[ |{\cal C}(\mathbf{v})| \geq  N \ | \ {\cal F}_L  \right] \leq n^{-\frac{1}{2} \ln d}.
\] 
Applying Theorem \ref{thrm:AcCL-Tail} for $l=1$ and $p=q=1$ and $\delta=1$, we have that
\begin{equation}\label{eq:BoundOnJBoundMarginal}
\Pr[\mathbf{v}(i) \geq 1 \ | \  {\cal F}_L ]  \leq  2\exp\left(-d^{7/10}\right) \qquad i=1, \ldots, \ell.
\end{equation}

\noindent
If $|{\cal C}(\mathbf{v})|\geq N$,  then there should be  a subset $D$ of vertices in $L$ such that 
$|D|=N$ and $D\subseteq {\cal C}(\mathbf{v})$. 
Applying a union bound gives
\begin{eqnarray}
\Pr\left[ |{\cal C}(\mathbf{v})| \geq  N \ | \ {\cal F}_L\right] & \leq & {\ell  \choose N}
\Pr[\mathbf{v}(i) \geq 1| {\cal E}_L ]  ^N  \hspace{2.25cm} \mbox{[independence of $\mathbf{v}(i)$s]}\nonumber \\
& \leq & {\frac{\ln n}{(\ln  d)^5}\choose N}
2^N\exp\left(-d^{7/10} \right)^N \hspace{1.5cm} \mbox{[we use \eqref{eq:BoundOnJBoundMarginal} ]}
\nonumber \\
&\leq&  n^{\frac{\ln 2}{(\ln  d)^5}}2^N\exp\left(- d^{7/10}N \right)
\hspace*{1.775cm} \left[ \textrm{as }{\ell \choose N}\leq 2^{\ell} \right]\nonumber \\
&\leq& n^{2\frac{\ln 2}{(\ln d)^5}}\cdot n^{- \ln d}
\leq n^{-\frac12\ln d}.
\nonumber 
\end{eqnarray}
The lemma follows.

\subsection{Proof of Lemma \ref{lemma:exptRj-bounding}.}\label{sec:lemma:exptRj-bounding}

When there is no danger of confusion, we abreviate ${\cal Q}_j(\mathbf{v})$ to ${\cal Q}_{j}$. It holds that
\begin{eqnarray} 
 \mathbb{E}\left[e^{t|{\cal Q}_j|} \ | \  {\cal F}_L \right]&=&
\sum_{r=0}^{\ell}e^{tr}\cdot \textrm{P}_{ {\cal F} }[|{\cal Q}_j|=r] 
\leq \sum_{r=0}^{\ell}e^{tr}\cdot \textrm{P}_{ {\cal F} }[|{\cal Q}_j |\geq r]
\nonumber\\
&\leq& \sum_{r=0}^{\ell }e^{tr}\ 
\max_{q} \left \{  \textrm{P}_{ {\cal F} } \left [ \prod^{r+q}_{s=q} \mathbf{v}(s)\geq 1  \ | \ \mathbf{v}(q)\geq 1 \right] \right\}.
\label{eq:R1MgfBound}
\end{eqnarray}
So as to deal with the conditional probability  in \eqref{eq:R1MgfBound} we work as follows:
It holds that
\begin{equation}\label{eq:R1MgfBoundRelvCond}
\textrm{P}_{ {\cal F} } \left [ \prod^{r+q}_{s=q} \mathbf{v}(s)\geq 1  \ | \ \mathbf{v}(q)\geq 1 \right] \leq 
\frac{\textrm{P}_{ {\cal F} } \left [ \prod^{r+q}_{s=q} \mathbf{v}(s)\geq 1  \right]}{\textrm{P}_{\cal F}[\mathbf{v}(q)\geq 1]}.
\end{equation}

\noindent
Consider $\G$ and the vertex $w_q \in L$. Then,  it holds that
\begin{eqnarray}
\Pr[\mathbf{v}(q)\geq 1\  | \ {\cal F}_L] &\geq&  \Pr[ {\tt degree}(w_q)\geq (1+\alpha)d \ |\ {\cal F}_L] \nonumber \\
&\geq &
\Pr[ {\tt degree}(w_q)=(1+\alpha)d \ |\ {\cal F}_L]\geq  \exp(-d),  \label{eq:ClDeNominatorBound}
\end{eqnarray}
where the last derivation follows from Stirling's approximation.

We use the tail bound we obtained in Theorem \ref{thrm:AcCL-Tail} and get that 
\begin{equation}
\textrm{P}_{ {\cal F} } \left [ \prod^{r+q}_{s=q} \mathbf{v}(s)\geq 1  \right]  \leq 4\exp\left[-d^{7/10} (r+1)\right].
\label{eq:ClNominatorBound}
\end{equation}
Plugging into \eqref{eq:R1MgfBoundRelvCond} the bounds from \eqref{eq:ClNominatorBound} and
\eqref{eq:ClDeNominatorBound} we get that
\[
\textrm{P}_{ {\cal F} } \left [ \prod^{r+q}_{s=q} \mathbf{v}(s)\geq 1  \ | \ \mathbf{v}(q)\geq 1 \right]  \leq \exp\left(d-d^{7/10}(r+1) \right).
\]

\noindent
In turn, plugging the above into (\ref{eq:R1MgfBound}) we have that
\begin{eqnarray}
 \mathbb{E}\left[e^{t|{\cal Q}_j|} \ |\  {\cal F}_L \right]
&\leq&  \exp(d-d^{7/10}) \sum_{r=0}^{\ell  }\exp\left[(t-d^{7/10}) r\right]
\ \leq  \  \exp(d-d^{-7/10}) \sum_{r=0}^{\ell }\exp\left(-0.8d^{7/10}r\right) 
\nonumber \\
&\leq&  \exp(d-d^{7/10})\left({1-\exp\left( -0.8d^{7/10}\right)} \right)^{-1} \ \leq \ \exp(d),
\nonumber 
\end{eqnarray}
in the last derivation of the first line we use the fact that $0\leq t\leq d^{3/5}$.
The lemma follows.

\section{Proof of Theorem \ref{thrm:CL-Tail}}\label{sec:thrm:CL-Tail}

\noindent
For the sake of brevity let $\WeightPathA (P)=\prod^{\ell}_{i=1} W(v_i)$.
For every $u\in P$, let  $\mathbf{Y}_u$ be the event that ``vertex $u$ has less than 
$2d^{9/10}$ neighbours in $P$." 
Also, let  $\MaxDegEvent=\cap_{u\in P}\mathbf{Y}_u$.
For  any  $t>0$ we have that 
\begin{eqnarray}
 \Pr[\WeightPathA (P) \geq \delta \ |\  {\cal E}_{ P}] &\leq &\Pr[{\MaxDegEvent}^c \ | \ {\cal E}_{ P}]+
 \Pr[ \WeightPathA (P)\geq \delta \ |\  \MaxDegEvent, {\cal E}_{ P} ] \nonumber\\
 &\leq &\Pr[{\MaxDegEvent}^c \ | \ {\cal E}_{ P} ]+ \Pr[\WeightPathA ^t(P) \geq \delta^t \ |\ \MaxDegEvent, {\cal E}_{ P}] \nonumber \\
  &\leq &\Pr[\MaxDegEvent^c\ |\  {\cal E}_{ P}]+ \delta^{-t}\  
  \mathbb{E}\left[\WeightPathA ^t(P) \ |\ \MaxDegEvent, {\cal E}_{ P} \right], 
  \label{eq:CLTailBase}
\end{eqnarray}
 where the last inequality follows from Markov's inequality.  
 The theorem follows by bounding appropriately the quantities 
 on the r.h.s. of  \eqref{eq:CLTailBase}.
When there is danger of confusion we abbreviate $\mathbb{E}[\cdot \ |\ \MaxDegEvent, {\cal E}] $ 
and $\Pr [\cdot| \ \MaxDegEvent, {\cal E}_P]$ to $\mathbb{E}_{ \MaxDegEvent, {\cal E}} [\cdot]$
and $\textrm{P}_{ \MaxDegEvent, {\cal E}} [\cdot]$, respectively.

For any $v\in P$ let $\mathbf{Y}^c_v$  be the complement of the event $\mathbf{Y}_v$. We have that 
\begin{eqnarray}
\Pr[\mathbf{Y}^c_v\  |\ {\cal E}_P] & \leq & \sum_{j\geq 2d^{9/10}} {\ell \choose j}\left( \frac{d}{n}\right)^j\left( 1-\frac{d}{n}\right)^{\ell-j} 
\; \leq \; \sum_{j\geq 2d^{9/10}}\left( \frac{\ell e d}{jn} \right)^j 
\hspace{1.4cm} \left [\textrm{as ${\ell \choose j}\leq (\ell e/j)^j$} \right]\nonumber \\
&\leq & \sum_{j\geq 2d^{9/10}}\left( \frac{50 e d^{1/10}\ln n}{n} \right)^j \; \leq \;2\left( \frac{100 e d^{1/10}\ln n}{ n} \right)^{2d^{9/10}}
\ \leq \ n^{-d^{9/10}}.
\label{eq:ProbHvBound}
\end{eqnarray}
Noting  that $\Pr[\MaxDegEvent^c\ | \ {\cal E}_{ P}] = \Pr[\cup_{v\in P} \mathbf{Y}^c_v\ | \ {\cal E}_{ P} ]$ a
simple union bound yields
\begin{eqnarray}\label{eq:PrACBound}
 \Pr[\MaxDegEvent^c \ |\  {\cal E}_{ P} ] \leq \sum_{v\in P} \Pr[\mathbf{Y}^c_v \ |\  {\cal E}_P]
\ \leq \ n^{1-d^{9/10}} ,
\end{eqnarray}
where in the second inequality we use \eqref{eq:ProbHvBound}.
For every $v\in P$ we  define  the new weight  $\bar{ W}(v)$ such that 
\begin{equation}\nonumber
 \bar{W}(v)=\left \{
\begin{array}{lcl}
(1+\gamma)^{-1} &\qquad& \textrm{if ${\tt degree}(v)\leq (1+ \frac{9}{10}\alpha)d$}\\ 
d^{(2c)}\cdot {\tt degree}(v) && \textrm{otherwise}.
\end{array}
\right .
\end{equation}
where $\alpha, \gamma,c$ are defined in the statement of Theorem \ref{thrm:CL-Tail}.
It is elementary to verify that with the new weights we have 
\[
\WeightPathA(P)\leq \prod_{v\in P}\bar{W}(v).
\]
Furthermore, we have the following result.
\begin{lemma}\label{lemma:factorEexpC(L)}
Let $  \alpha,\gamma,c, d, \ell$  and  $P$ be as in Theorem \ref{thrm:CL-Tail}.
For  any $t>0$, it holds that
\[
\mathbb{E}_{ \MaxDegEvent, {\cal E}}  \left[ \WeightPathA^t(P)  \right] \leq
 \prod_{v\in P}  \mathbb{E} _{ \MaxDegEvent, {\cal E}}  \left[ \bar{W}^t(v)  \right ].
\]
\end{lemma}
The proof of Lemma \ref{lemma:factorEexpC(L)} appears in Section 
\ref{sec:lemma:factorEexpC(L)}.

\begin{lemma}\label{lemma:BoundEexpRv}
Let  $\alpha,\gamma,c, d, \ell$  and  $P$ be as in Theorem \ref{thrm:CL-Tail}.
 For any integer  $t$ such that $0<t< \gamma^{-3} \  d^{4/5}$ and any 
$v \in P$ we have that
\[
\mathbb{E}_{ \MaxDegEvent, {\cal E}}  \left[ \bar{W}^t(v) \right]\leq 
 (1+\gamma)^{-t}+  \exp\left(-\alpha^2d/10\right).
\]
\end{lemma}
The proof of Lemma  \ref{lemma:BoundEexpRv} appears in Section \ref{sec:prop:BoundEexpRv}.

From Lemma \ref{lemma:factorEexpC(L)} and Lemma  \ref{lemma:BoundEexpRv}
we get that
\begin{eqnarray}
\mathbb{E}_{ \MaxDegEvent, {\cal E}} \left[ \WeightPathA ^t(P) \right] &\leq&
\left[(1+\gamma)^{-t} + \exp\left(-\alpha^2d/10\right) \right]^{\ell} 
\nonumber \\
&\leq& (1+\gamma)^{-t\ell}
\left[1+\exp\left(-\alpha^2d/10 +\gamma t\right) \right]^{\ell }  \hspace{1.75cm} \mbox{[as $(1+\gamma)^t\leq e^{\gamma t}$]}
\nonumber\\
&\leq& \exp\left(- \frac{\gamma}{1+\gamma} t \ell  + \exp\left(-\alpha^2 5d/10+ \gamma t\right) \ell  \right)
\hspace{.9cm} \mbox{[as $(1+\gamma)^{-t}\leq e^{-\frac{\gamma}{1+\gamma} t}$]}
\nonumber
\end{eqnarray}
For $T=5\frac{1+\gamma}{\gamma}d^{4/5}$,   we have that
\begin{equation}
\mathbb{E} _{ \MaxDegEvent, {\cal E}} \left[ \WeightPathA^{T}(P)  \right]  \leq \exp\left(-d^{4/5}\ell  \right). 
\label{eq:ECPGenerating}
\end{equation}
Setting $t=T$ in \eqref{eq:CLTailBase} and   plugging  \eqref{eq:PrACBound}, \eqref{eq:ECPGenerating}  yields
\begin{eqnarray}
 \Pr[\WeightPathA (P) \geq \delta \ |\  {\cal E}_{ P}]  & \leq  & n^{1-d^{9/10}} +\delta^{-T}\exp\left(-d^{4/5}\ell  \right) \nonumber \\
 &\leq & n^{1-d^{9/10}} +\exp\left(-d^{4/5} (\ell+\ln \delta)  \right) 
 \hspace{1.3cm}\mbox{[as  $\delta^{-T}\leq \exp(-d^{4/5}\ln \delta)$]}\nonumber \\
 &\leq & 2\exp\left(-d^{4/5} (\ell+\ln \delta)  \right).  \hspace{2.75cm}\mbox{[as $\ell\leq 100\ln n$ and $\delta>0$ is fixed]} \nonumber
\end{eqnarray}
The theorem follows.

\subsection{Proof of Lemma \ref{lemma:factorEexpC(L)}}\label{sec:lemma:factorEexpC(L)}

Consider $\G$ conditional on ${\cal E}_P$, i.e. $P$ forms a path in $\G$.

For each  $v_i \in P$,   let ${\tt deg}_{out}(v_i)$ be the number of vertices outside $P$ which are  adjacent to $v_i$ 
plus the  number of edges of $P$ which are incident to $v_i$. That is, we don't count the edges between $v_i$ 
and vertices $v_j\in $ such that $|i - j|>1$.

Let $\bar{W}_{out}(v)$ be the weight of vertex $v\in P$, defined as follows:
\begin{equation}\nonumber
 \bar{W}_{out}(v)=\left \{
\begin{array}{lcl}
(1+\gamma)^{-1} &\qquad& \textrm{if ${\tt deg}_{out}(v)\leq (1+ \frac{9}{10}\alpha)d$}\\ 
d^{2c}\cdot {\tt deg}_{out}(v) && \textrm{otherwise}.
\end{array}
\right .
\end{equation}
That is, $\bar{W}_{out}(v)$ differs with $\bar{W}(v)$ in that it considers only the ${\tt deg}_{out}(v)$,
whereas $\bar{W}(v)$ considers ${\tt degree}(v)$. Since ${\tt degree}(v)\geq {\tt deg}_{out}(v)$
it is elementary to show that 
\begin{equation}\label{eq:WoutVsWbar}
\bar{W}_{out}(v) \leq \bar{W}(v) \qquad \forall v\in P.
\end{equation}
Furthermore, conditioning on $\MaxDegEvent$ it holds that  that 
\begin{equation}\label{eq:WvsWout}
W(v) \leq \bar{W}_{out}(v) \qquad \forall v\in P.
\end{equation}
To see why the above holds consider the following:  Since the event $\MaxDegEvent$ holds,   for a vertex $v\in P$ 
such that ${\tt deg}_{out}(v)\leq \left(1+\frac{9\alpha}{10}\right )d $ we have  $W(v) = \bar{W}_{out}(v)= (1+\gamma)^{-1}$. 
Note that if ${\tt deg}_{out}(v)\leq \left( 1+\frac{9\alpha}{10}\right)d$, then the event $\MaxDegEvent$ guarantees 
that ${\tt degree}(v)\leq (1+\alpha)d$. Similarly, we have $W(v) = \bar{W}_{out}(v)$ for a vertex  $v$ such that 
${\tt deg}_{out}(v)>(1+\alpha)d$.

Furthermore, if   $v$ is such that $ (1+\frac{9}{10}\alpha)d<{\tt deg}_{out}(v) \leq (1+\alpha)d -2d^{9/10}$ 
we have  $W(v)=(1+\gamma)^{-1}<1$ while $\bar{W}_{out}(v)\gg 1$.  That is, 
 $W(v) < \bar{W}_{out}(v)$.

Finally,   consider the case    $ (1+\alpha)d -2d^{9/10} < {\tt deg}_{out}(v) \leq (1+\alpha)d$.
If, furthermore, we have ${\tt degree}(v)\leq (1+\alpha)$, then it is direct that   $W(v) < \bar{W}_{out}(v)$.
On the other hand, if ${\tt degree}(v)> (1+\alpha)$, then we show that we  have
$W(v) < \bar{W}_{out}(v)$.  For this, consider the ratio $W(v)/\bar{W}_{out}(v)$. It holds that
\[
\frac{W(v)}{\bar{W}_{out}(v)} \leq \frac{d^c({\tt deg}_{out}(v) + 2d^{9/10})}{d^{2c}{\tt deg}_{out}(v) } \leq
2d^{-c} < 1.
\]
For the one prior to last inequality we used the assumption that $ (1+\alpha)d -2d^{9/10} < {\tt deg}_{out}(v)$.
From the above we verify that \eqref{eq:WvsWout} is indeed true.
Then, we have that
\begin{eqnarray}
\mathbb{E}_{ \MaxDegEvent, {\cal E}}  \left[ \WeightPathA^t(P)   \right]  &= &
\mathbb{E}_{ \MaxDegEvent, {\cal E}}  \left[\prod_{v\in P}  W^t(v)   \right] 
\; \leq \; \mathbb{E}_{ \MaxDegEvent, {\cal E}}  \left[\prod_{v\in P} \bar{W}^t_{out}(v)   \right] 
\qquad \mbox{[ due to \eqref{eq:WvsWout} and $t>0$ ]} \nonumber \\
&\leq & \prod_{v\in P}  \mathbb{E}_{ \MaxDegEvent, {\cal E}}  \left[\bar{W}^t_{out}(v)   \right] \nonumber \\
&\leq & \prod_{v\in P}  \mathbb{E}_{ \MaxDegEvent, {\cal E}}  \left[\bar{W}^t(v)   \right] .
\hspace{5cm} \mbox{[ due to \eqref{eq:WoutVsWbar}]} \nonumber
\end{eqnarray}
The derivation in the second line follows from the observation that 
 the variables $\bar{W}_{out}(v_1), 
 \ldots, \bar{W}_{out}(v_{\ell})$ are 
 independent with each other.
The lemma follows.

\subsection{Proof of Lemma \ref{lemma:BoundEexpRv}}\label{sec:prop:BoundEexpRv}

\begin{lemmaproof}{\ref{lemma:BoundEexpRv}}
 So as to prove the lemma we need the following claim.

\begin{claim}\label{claim:WtBoundTech}
Let $d, \alpha, \gamma, t$ be as in the statement of  Lemma \ref{lemma:BoundEexpRv}.
For any $q$ such that $|q|  \leq \gamma^{-3}\ d^{4/5}$, 
the following is true: 
\[
 \sum^n_{i=(1+0.9\alpha)d+q}(i+2)^t{n \choose i}\left(\frac{d}{n}\right)^i\left(1-\frac{d}{n}\right)^{n-i}
\leq d^t\cdot \exp\left(-\alpha^2d/7\right).
\]
\end{claim}

\noindent
The proof of Claim \ref{claim:WtBoundTech} appears after this proof.

Let ${\tt deg}_{ex}(v)$ be the number of edges which are incident  to $v$, excluding those
which belong to $P$.  We have that
\begin{eqnarray}
\lefteqn{
\mathbb{E}_{\MaxDegEvent, {\cal E}} [\bar{W}^t(v) ]
}\hspace{.2in} \nonumber \\
 &=&(1+\gamma)^{-t} \textrm{P}_{\MaxDegEvent, {\cal E}} [{\tt deg}_{ex}(v) \leq (1+0.9\alpha)d-1 ]+
d^{2ct}\sum^n_{i= (1+0.9\alpha)d-2} (i+2)^t \textrm{P}_{\MaxDegEvent, {\cal E}} [{\tt deg}_{ex} (v)=i  ]
\nonumber \\
&\leq& (1+\gamma)^{-t}+ d^{2ct} 
\sum^n_{i=(1+0.9\alpha)d-2} (i+2)^t {n \choose i}\left(\frac{d}{n}\right)^i\left(1-\frac{d}{n}\right)^{n-i} 
\nonumber  \\ 
&\leq& (1+\gamma)^{-t}+ d^{t(2c+1)}  \exp\left(-\alpha^2d/7\right).
\hspace*{5.00cm} \mbox{[from Claim \ref{claim:WtBoundTech}]}
\nonumber\\
&\leq& (1+\gamma)^{-t}+   \exp\left(-\alpha^2d/9 \right).
\nonumber
\end{eqnarray}
the last inequality follows from the fact that $t<\gamma^{-3}\ d^{4/5}$ and $c$ is  fixed.
The lemma follows. 
\end{lemmaproof}
\vspace{.3cm}

\begin{claimproof}{\ref{claim:WtBoundTech}}
Recall that 
${n \choose i}=\frac{n}{i}{n-1\choose i-1}$, for $1\leq i\leq n$. Using this equality we have that
\begin{eqnarray}
\lefteqn{
\sum^n_{i=(1+0.9\alpha)d+q}(i+2)^t{n \choose i}\left(\frac{d}{n}\right)^i\left(1-\frac{d}{n}\right)^{n-i}
}\hspace{.2in}\nonumber \\
&=& d\sum^{n}_{i=(1+\alpha)d+q}\left(1+\frac2i\right)^ti^{t-1}{n -1 \choose i-1}\left(\frac{d}{n}\right)^{i-1}\left(1-\frac{d}{n}\right)^{n-1-(i-1)}
\nonumber \\
&\leq&d\sum^{n-1}_{j=(1+\alpha)d+q-1}\left(1+\frac2{j+1}\right)^t (j+1)^{t-1}{n -1 \choose j}\left(\frac{d}{n}\right)^{j}\left(1-\frac{d}{n}\right)^{n-1-j}.
\nonumber \qquad \mbox{[we set $j=i-1$]}
\end{eqnarray}
It is direct  that repeating the exactly the same calculation $l$ times, where $l\leq t$, we get that
\begin{eqnarray}
\lefteqn{
\sum^n_{i=(1+0.9\alpha)d+q}(i+2)^t{n \choose i}\left(\frac{d}{n}\right)^i\left(1-\frac{d}{n}\right)^{n-i}
}\hspace{.2in}\nonumber \\
 &\leq& d^l\left(\prod_{s=0}^{l-1}\left(1+\frac{s}{d}\right)\right)\ 
\sum_{j=(1+0.9\alpha)d+q-l}^{n-l}\left(1+\frac2{j+l}\right)^t(j+l)^{t-l} {n-l \choose j}\left(\frac{d}{n}\right)^{j}\left(1-\frac{d}{n}\right)^{n-l-j}.
\nonumber
\end{eqnarray}
In particular,   for $l=t$  the above inequality implies the following
\begin{eqnarray}
\lefteqn{
\sum^n_{i=(1+0.9\alpha)d+q}(i+2)^t{n \choose i}\left(\frac{d}{n}\right)^i\left(1-\frac{d}{n}\right)^{n-i}
}\hspace{1in}\nonumber \\
 &\leq & d^t \  e^{2} \  \Pr[{\cal B}(n-t,d/n)>(1+0.9\alpha)d+q-t] \  \prod_{s=0}^{t-1}\left(1+\frac{s}{d}\right),
 \label{eq:SBoundWrtT}
 \end{eqnarray}
where the probability term expresses the probability that a random variable
distributed as in binomial distribution with parameters $n-t$ and $d/n$
is at least $(1+\alpha)d+q-t$. 

Using the fact that $|q|,|t|<\gamma^{-3}\ d^{4/5}$ and  standard large deviation results about the binomial distribution,
i.e. Corollary 2.3 from \cite{luczak} we get that 
\begin{eqnarray}\label{eq:Cor2.3luczak}
\Pr[{\cal B}(n-t, d/n)\geq (1+0.9\alpha)d+q-t] \leq \exp\left(-\alpha^2d/6\right). \label{eq:Bin(n-t, d/n)tail}
\end{eqnarray}
Also, we have that 
\[
\prod_{s=0}^{t-1}\left(1+\frac{s}{d}\right)\leq \exp\left(\sum_{s=0}^{t-1}s/d\right)
\leq \exp\left[  t(t-1)/(2d)\right] \leq \exp\left(  \gamma^{-6}d^{3/5}/2\right). \qquad \mbox{[as $t\leq \gamma^{-3} \ d^{4/5}$ ]}
\]
The claim follows by plugging the above and \eqref{eq:Cor2.3luczak} into  \eqref{eq:SBoundWrtT}.
\end{claimproof}

\section{Poof of Theorem \ref{thrm:AcCL-Tail}}\label{sec:thrm:AcCL-Tail}

For the sake of brevity, let 
$
\WeightPathB= \WeightPathB (L, r, q)=\prod^{r+q-1}_{j=q}\mathbf{v}(j).
$
Consider $\mathbf{v}$ which is distributed as in $\zeta$.  Let  $\MaxDegEvent$ be the event that 
`` the maximum degree of  $\G$ that induces $\mathbf{v}(i)$ is at most $ (\ln n)^2$, for $i=1, \ldots, l$''. 
For any $t>0$, we have that
\begin{eqnarray}
\Pr[ \WeightPathB  \geq \delta \; |\; {\cal F}_L ] &\leq & \Pr[\MaxDegEvent^c \; | \; {\cal F}_L]+
\Pr[ \WeightPathB  \geq \delta \; |\; {\cal F}_L,   \MaxDegEvent] \nonumber\\
&\leq &\Pr[\MaxDegEvent^c \; | \; {\cal F}_L]+
\Pr[ \WeightPathB^t   \geq \delta^t \; |\; {\cal F}_L,   \MaxDegEvent]  \nonumber \\
&\leq &\Pr[\MaxDegEvent^c \; | \; {\cal F}_L]+
\delta^{-t}\ {\mathbb{E}[ \WeightPathB^t   \; |\; {\cal F}_L,   \MaxDegEvent]} 
\label{eq:AcCL-TailDegBound}
\end{eqnarray}
where that last derivation follows from Markov's inequality.
The theorem follows by bounding appropriately the terms in the r.h.s. of \eqref{eq:AcCL-TailDegBound}.

As far as $\Pr[\MaxDegEvent^c \; | \; {\cal E}_L]$ is concerned,  we have the following claim.

\begin{claim}\label{claim:MaxDegreeEp}
It holds that
$ \Pr[ \MaxDegEvent^c \; | \; {\cal F}_L ] \leq  \exp\left( -(1/2)(\ln n)^2\right).$
\end{claim}
\begin{proof}
Consider an instance of $\G$ and conditional that $L$ is a path in the graph.

An internal  vertex in $L$ gets degree $(\ln n)^2$ if, apart from  its two neighbours in the path there are at least 
another $(\ln n)^2-2$ vertices to get connected. Applying, standard Chernoff's bound we have that this happens with probability
at most $\exp(-(3/4)(\ln n)^2)$. Similarly,  each of $w_1$ and $w_{l}$, the ends of $L$, get degree 
$(\ln n)^2$ if, apart from the its two neighbours in the path there are another $(\ln n)^2-1$ to get connected.
Chernoff's bound implies that this happens with probability at most $\exp(-(3/4)(\ln n)^2)$. Finally,  each vertex
$u$ not in $L$, gets degree $(\ln n)^2$ with probability at most $\exp(-(\ln n)^2)$. This result follows from Chernoff's 
bound, too.

The above argument and a simple union bound implies that the probability for $\G$, conditional on
that $L$ appears, to have maximum degree greater than $(\ln n)^2$ is $n\exp(-\frac{3}{4}(\ln n)^n)$.

For the probability term $ \Pr[ \MaxDegEvent^c \; | \; {\cal F}_L ]$, instead of one copy of $\G$
conditional the path $L$, we consider $l$ independent copies.  
The claim follows by taking a union bound and noting that $l=\Theta(\ln n)$. 
\end{proof}

\noindent
For the sake of brevity we let $\mathbb{E}_{ \MaxDegEvent, {\cal F}} [\cdot]$ stand for the operator
$\mathbb{E}[\cdot| \MaxDegEvent, {\cal F}_L]$. Also, let $\textrm{P}_{ \MaxDegEvent, {\cal F}} [\cdot]$
stand for  $\Pr [\cdot| \MaxDegEvent, {\cal F}_L]$.  We have that

\begin{proposition}\label{prop:BoundEexpFv}
Let  $ \alpha,\gamma,c, d,  l $ and $L$ are as in the statement of Theorem \ref{thrm:AcCL-Tail}.
For any   $0<t<(\gamma^{-3})\ d^{7/10}$  the following it true:  
%
For any $j=1,\ldots, l$ we have that
\[
\mathbb{E} _{ \MaxDegEvent, {\cal F}}\left[ \mathbf{v}^t(j)  \right]\leq 
(1+\gamma)^{-t}  \left(1+2\exp\left(-{d^{4/5}}/{2}\right)\right).
\]
\end{proposition}
The proof of Proposition \ref{prop:BoundEexpFv} appears in Section
\ref{sec:prop:BoundEexpFv}.

From  Proposition \ref{prop:BoundEexpFv} we get that
\begin{eqnarray}
\mathbb{E} _{ \MaxDegEvent, {\cal F}}[ \WeightPathB^t (P) ]&\leq&
(1+\gamma)^{-t r }\left(1+
2\exp\left(-{d^{4/5}}/{2}\right)  \right)^{r},
\nonumber \\ 
&\leq & \exp \left( -\frac{\gamma}{1+\gamma} t r+2\exp\left(-{d^{4/5}}/{2}\right)r \right),
\end{eqnarray}
where  used the inequality $1+x\leq e^{x}$.
For $T=5\frac{1+\gamma}{\gamma}d^{7/10}$ we have that 
\begin{eqnarray}
\mathbb{E} _{ \MaxDegEvent, {\cal F}}[ \WeightPathB ^{T} ]\leq \exp\left(-d^{7/10}r  \right). \label{eq:EXPTAK(P)}
\end{eqnarray} 
Setting $t=T$ in  \eqref{eq:AcCL-TailDegBound} and plugging  \eqref{eq:EXPTAK(P)} and Claim
\ref{claim:MaxDegreeEp} gives
\begin{eqnarray}
\Pr[ \WeightPathB  \geq \delta \; |\; {\cal F}_L ] &\leq &n^{-\frac{3}{4}\ln n+1}+\delta^{-T}\exp\left(-d^{7/10}r  \right) 
\ \leq \ \exp\left(-d^{7/10}(r+\ln \delta)  \right), \nonumber
\end{eqnarray}
where the last derivation follows from noting that  $\delta^{-T} \leq \exp(-d^{7/10} \ln \delta)$, for fixed $\delta>0$, 
and $r=O(\ln n)$.
The theorem follows.

\subsection{Proof of Proposition \ref{prop:BoundEexpFv}}\label{sec:prop:BoundEexpFv}

\newcommand\OutPathDeg{{\tt deg}_{out}}
\newcommand\MaxDegEventB{ \mathbold{A} }


  For $w_i \in L$ in $\G$, let $\OutPathDeg(w_1)$ denote the number of edges between $v$ and the vertices outside 
 $L$.  
%
%
By definition it holds that
\begin{eqnarray} 
\lefteqn{
 \mathbb{E} _{ \MaxDegEventB, {\cal F}} [\mathbf{v}^t(j) ] \ \leq \ 
\sum_{i=0}^{ (\ln n)^2}\mathbb{E} _{ \MaxDegEventB, {\cal F}} [\mathbf{v}^t(j)  \  | \  \OutPathDeg(w_j)=i] 
\ \textrm{P} _{ \MaxDegEventB, {\cal F}}[\OutPathDeg(w_j)=i ] 
 }  \hspace{.2in} \nonumber \\
 &\leq& \sum_{i=0}^{(\ln n)^2} \mathbb{E} _{ \MaxDegEventB, {\cal F} }   [W^t(w_j) \ |\  \OutPathDeg(w_j)=i]\ 
\mathbb{E} _{ \MaxDegEventB, {\cal F}}[Q^t(w_j) \ | \  \OutPathDeg(w_j)=i]\  
\textrm{P} _{ \MaxDegEventB, {\cal F}}[\OutPathDeg(w_j)=i ].\qquad \label{eq:U^tBase}
\end{eqnarray}
The second deviation follows by  observing that,   conditional on $\OutPathDeg(w_j)$,  the  variables 
$W(w_j)$ and $Q(w_j)$  are independent with each other. 
The proposition follows by  bounding  the  r.h.s. of  \eqref{eq:U^tBase}.

\begin{lemma}\label{lemma:SplitF-MGF}
Let $\alpha,\gamma,c, d,  t, \ell$ and $L$ be as in the statement of Proposition \ref{prop:BoundEexpFv}. For any
$j\in [\ell]$ it holds that 
\begin{equation}
\mathbb{E} _{ \MaxDegEventB, {\cal F}} [ Q^t(w_j) \  | \   \OutPathDeg(w_j) ]<1+2d\exp\left(-d^{4/5}\right)  \OutPathDeg(w_j).
\label{eq:ExptQ^tGivenDegree}
\end{equation}
\end{lemma}
The proof of Lemma \ref{lemma:SplitF-MGF} appears in Section
\ref{sec:lemma:SplitF-MGF}.

Plugging the bound from Lemma \ref{lemma:SplitF-MGF} into  into \eqref{eq:U^tBase},  yields
\begin{eqnarray}
\lefteqn{
 \mathbb{E} _{ \MaxDegEventB, {\cal F}} [\mathbf{v}^t(j)    ] 
  }  \hspace{.2in} 
 \nonumber \\
 &\leq &\mathbb{E} _{ \MaxDegEventB, {\cal F}} [W^t(w_j)   ]+2d\exp\left(-d^{4/5}\right)
\sum_{i=0}^{(\ln n)^2} i  \  \mathbb{E}  _{ \MaxDegEventB, {\cal F}}[W(w_j)   \ | \  \OutPathDeg(w_j)=i ]\ 
\textrm{P} _{ \MaxDegEventB, {\cal F}}[\OutPathDeg(w_j)=i  ].\qquad\label{eq:EUt-Base}
\end{eqnarray}

\begin{lemma}\label{lemma:beautifyMGF-U}
Let  $\alpha,\gamma,c, d,t, \ell$ and $L$ be as in the statement of Proposition \ref{prop:BoundEexpFv}.
For any $j\in [\ell]$ it holds that
\[
 \sum_{i=0}^{ (\ln n)^2  }i  \  \mathbb{E} _{ \MaxDegEventB, {\cal F}}[W^t(w_j) \ |\  \OutPathDeg(w_j)=i]
\  {\rm P} _{ \MaxDegEventB, {\cal F}}[\OutPathDeg(w_j)=i  ]
\leq 2d{(1+\gamma)^{-t}}+ \exp\left(-\alpha^2d/6\right). \nonumber
\]
\end{lemma}
The proof of Lemma \ref{lemma:beautifyMGF-U} appears in Section \ref{sec:lemma:beautifyMGF-U}.

\begin{lemma}\label{lemma:EWtInt0Repeat}
Let  $\alpha,\gamma,c, d,t ,\ell$ and $L$ be as in the statement of Proposition \ref{prop:BoundEexpFv}.
For any $j\in [\ell]$ it holds that
\[
\mathbb{E} _{ \MaxDegEventB, {\cal F}}[W^t(w_j)  ]\leq
(1+\gamma)^{-t}+  \exp\left(-\alpha^2d/10\right).
\]
\end{lemma}
We omit the proof of Lemma \ref{lemma:EWtInt0Repeat} as it 
is identical to that  of Lemma \ref{lemma:BoundEexpRv}  (see Section \ref{sec:prop:BoundEexpRv}).

Plugging into (\ref{eq:EUt-Base}) the bounds from Lemmas \ref{lemma:beautifyMGF-U},
\ref{lemma:EWtInt0Repeat} we get that
\begin{eqnarray}
 \mathbb{E} _{ \MaxDegEventB, {\cal F}} [\mathbf{v}^t(j) ]&\leq& 
 \left[ (1+\gamma)^{-t} + \exp\left(-\alpha^2d/10\right) \right]
\left(1+\exp\left(-{d^{4/5}}/{2}\right)\right)
 \nonumber \\
 &\leq& (1+\gamma)^{-t}
 \left(1+2\exp\left(-{d^{4/5}}/{2}\right)\right).
\hspace{3.2cm} \mbox{[as $t\leq \gamma^{-3}d^{7/10}$]}\nonumber
\end{eqnarray}
The proposition follows.

\subsection{Proof of Lemma \ref{lemma:SplitF-MGF}}\label{sec:lemma:SplitF-MGF}

\begin{lemmaproof}{\ref{lemma:SplitF-MGF}}
For $\OutPathDeg(w_j)=0$ it is direct that $Q(w_j)=1$. This implies that   \eqref{eq:ExptQ^tGivenDegree} is true 
 for  $\OutPathDeg(w_j)=0$. In what follows  we assume that $\OutPathDeg(w_j)>0$.

Let $u_1,\ldots,u_{(\OutPathDeg(w_j))}$ be the neighbours of $w_j$ outside  $L$.
Also, for each $u_i$ let $S_i$ be the set of paths of length $\ln n/d^{4/5}$ 
that start from $u_i$ but do not use the vertices in $L$. For $1\leq i\leq \OutPathDeg(w_j)$,
let ${\cal E}_{i,x}$ be the event that there is a path $P \in S_j$ such that
 $\prod_{u\in P}W(u) \geq x$. 

Using Theorem \ref{thrm:CL-Tail} we get the following result whose proof follows after this proof.
\begin{claim}\label{claim:E1xTail}                        
For $1\leq i\leq \OutPathDeg(w_j)$ and any fixed  $x\geq 1$, it holds that
\[
{\rm  P} _{ \MaxDegEventB, {\cal F}} [{\cal E}_{i,x}]\leq 2d\  \exp\left(-d^{4/5}(\ln x+1)\right). 
\]
\end{claim}

\noindent
We have that 
\begin{eqnarray}
\lefteqn{
 \mathbb{E}_{ \MaxDegEventB, {\cal F}} [Q^t(w_j) \ | \ \OutPathDeg(w_j)]
 } \hspace{.2in}  \nonumber \\
 &=& \textrm{P} _{ \MaxDegEventB, {\cal F}} [Q(w_j)< 1\ |\ \OutPathDeg(w_j)] +
\mathbb{E} _{ \MaxDegEventB, {\cal F}} [Q^t(w_j)\ |\ \OutPathDeg(w_j),Q(w_j)\geq 1]\  
\textrm{P} _{ \MaxDegEventB, {\cal F}}[Q(w_j)\geq 1\ |\ \OutPathDeg(w_j)]
\nonumber \\
&\leq&1 + \mathbb{E}_{ \MaxDegEventB, {\cal F}}[Q^t(w_j)\ |\ \OutPathDeg(w_j),Q(w_1)\geq 1]\ 
\textrm{P} _{ \MaxDegEventB, {\cal F}} [Q(w_j)\geq 1\ |\ \OutPathDeg(w_j)]  
\nonumber \\
&\leq&1 + \mathbb{E}_{ \MaxDegEventB, {\cal F}}[Q^t(w_j)\mathbf{1}\{Q(w_j)\geq 1\}\  |\ \OutPathDeg(w_j)]
\label{eq:Q^tBase},
\end{eqnarray}
where in the second line we use the fact that $\textrm{P} _{ \MaxDegEventB, {\cal F}} [Q(w_j)< 1\ |\ \OutPathDeg(w_j)]\leq 1$.
 Also, it holds that
\begin{eqnarray} 
 \mathbb{E}_{ \MaxDegEventB, {\cal F}}[Q^t(w_j)\mathbf{1}\{Q(w_j)\geq 1\}\  |\ \OutPathDeg(w_j)]
&=& \int_{1}^{\exp( (\ln n)^4)} x^{t} \ 
\textrm{P} _{ \MaxDegEventB, {\cal F}} [Q(w_j) = x\ |\ \OutPathDeg(w_j) ]dx
\nonumber\\
&\leq& \int_{1}^{\exp(( \ln n)^4)} x^{t} \ \textrm{P} _{ \MaxDegEventB, {\cal F}} [Q(w_j)\geq x \ |\ \OutPathDeg (w_j)]dx.
\nonumber
\end{eqnarray}
The bound $\exp( (\ln n)^4 )$ follows from a simple calculation which suggests
that, given that the maximum degree is $ (\ln n)^2$, for any path $P \in S_j$
it holds that $\prod_{u\in P}W(u) \leq \exp\left( (\ln n)^4 \right)$. Note that
\begin{eqnarray}
 \textrm{P} _{ \MaxDegEventB, {\cal F}} \left [Q(w_j)\geq x\ |\ \OutPathDeg(w_j) \right]&=& 
 \textrm{P} _{ \MaxDegEventB, {\cal F}} \left [\cup_{i}{\cal E}_{i,x} \  | \  \OutPathDeg(w_j) \right]
\nonumber \\&\leq& 
 \OutPathDeg(w_j) \ \textrm{P} _{ \MaxDegEventB, {\cal F}} [{\cal E}_{1,x}] \hspace{3.64cm}\mbox{[from the union bound]}\nonumber\\
&\leq& 2d\exp\left(-d^{4/5}(1+\ln x)\right)\ \OutPathDeg(w_j). \hspace{1cm} \mbox{[from Claim \ref{claim:E1xTail}]}
\nonumber
\end{eqnarray}
We get that
\begin{eqnarray}
\lefteqn{
 \mathbb{E}_{ \MaxDegEventB, {\cal F}}[Q^t(w_j)\mathbf{1}\{Q(w_j)\geq 1\} \ |\ \OutPathDeg(w_j)]
 } \hspace{1in}  \nonumber \\
&\leq& 2d\exp(-d^{4/5})\OutPathDeg(w_j) \int_{1}^{\exp(( \ln n)^4)} x^{t}\exp\left(-d^{4/5}\ln x\right)dx
\nonumber \\
&\leq& 2d  \exp(-d^{4/5}) \OutPathDeg(w_j) \int_{1}^{\exp(( \ln n)^4)}\exp\left(-\frac{1}{2}d^{4/5}\ln x\right) dx
\hspace{1cm} \mbox{[as $0\leq t\leq \gamma^{-3}d^{7/10}$]}\nonumber\\
&\leq& 2d \exp(-d^{4/5}) \OutPathDeg(w_j). \nonumber
\end{eqnarray}
where in the final derivation we used the fact that $\int_1^{e^{ (( \ln n)^4)}}x^{-d^{4/5}/2}dx\leq 1$. 

The lemma follows by plugging the above bound  into (\ref{eq:Q^tBase}).
\end{lemmaproof} 
\vspace{.3cm}

\begin{claimproof}{\ref{claim:E1xTail}}
We use the same terminology as that in the proof of Lemma \ref{lemma:SplitF-MGF}.
That is,   we have  $u_1,\ldots, u_{(\OutPathDeg(w_j))}$   the neighbours of $w_j$ outside $L$. 
Also, for $u_i$ let $S_i$ be the set of paths, of length at most $\ln n/d^{4/5}$,  that start from $u_i$
but do not use  the vertices in $L$.

Note that the events ${\cal E}_{i,x}$ for $i=1,2,\ldots, \OutPathDeg(w_j)$ are symmetric. For this
reason we can focus only on the event ${\cal E}_{1,x}$.
${\cal E}_{1,x}$  occurs only if  there is a path $P\in S_{1}$ such that $\prod_{u\in P}W(u) \geq x$. 
For every $P\in S_1$ let $\mathbf{1}\{P\}$ be  an indicator variable which is $1$ if $\prod_{u\in P}W(u) \geq x$,
otherwise it is zero.

In what follows,  we l et $S^i_1\subseteq S_1$ denote the paths in $S_1$ of $i$ vertices.
 We have that
\begin{eqnarray}
\textrm{P}_{\MaxDegEventB, {\cal F} } [{\cal E}_{1,x}]  & \leq & \left( {\Pr[\MaxDegEventB\ | \  {\cal F}_L]} \right)^{-1} \ 
 \Pr[{\cal E}_{1,x} \ | \ {\cal F}_L] 
\; \leq \; (3/2)  \Pr \left[\sum_{P \in S_{1}} \mathbf{1}\{P\}  >0 \  | \  {\cal F}_L \right]
\qquad \mbox{[as $\Pr[\MaxDegEventB\ |\ {\cal F}_L]\geq 2/3$] } \nonumber \\
&\leq &  (3/2) \  \mathbb{E} \left[ \sum_{P \in S_{1}} \mathbf{1}\{P\} \ |\  {\cal F}_L \right],
\label{eq:ExTailBase}
\end{eqnarray}
where the last derivation uses Markov's inequality. The derivation
for the bound  $\Pr[\MaxDegEventB\ |\ {\cal F}_L]\geq 2/3$  are very similar to ones we have used before.

From Theorem \ref{thrm:CL-Tail} we get the following:   For any $P\in S^i_1$ we have that
\begin{equation}
\Pr [\mathbf{1}\{P \}=1 \ |\  {\cal F}_L, P\in S^i_1 ]\ \leq \exp\left[-d^{4/5}(i+\ln x)\right]. \label{eq:I_LMarginal}
\end{equation}
From  the linearity of expectation, we have that
\begin{eqnarray}
 \mathbb{E} \left[\sum_{P\in S_{1}} \mathbf{1}\{P\} \ |\  {\cal F}_L\right]&=&\sum_{i=0}^{\ln n/d^{4/5}}
{n \choose i}\left(\frac{d}{n}\right)^i \Pr[\mathbf{1}\{P\}=1| {\cal F}_L,  P\in S^{i}_1 ] \nonumber \\
&=&\sum_{i=0}^{\ln n/d^{4/5}} 
{n \choose i}\left(\frac{d}{n}\right)^i\exp\left(-d^{2/5}(i+\ln x)\right) 
\hspace{2.87cm}\mbox{[from (\ref{eq:I_LMarginal})]}\nonumber \\
&\leq&\exp(-d^{4/5}\ln x)\sum_{i=0}^{\ln n/d^{4/5}} \exp\left(-d^{4/5}i +i \ln d\right) 
\qquad \mbox{[as ${n\choose i}(d/n)^i\leq d^i$]}\nonumber \\
&\leq&\exp\left (-d^{4/5}\ln x-d^{4/5}+\ln d \right)\  \left( 1+e^{-d^{4/5}+\ln d} \right )^{-1}. \nonumber
\end{eqnarray}
The claim follows by noting that $(1+e^{-d^{4/5}+\ln d})^{-1}\leq 4/3$.
\end{claimproof}

\subsection{Proof of Lemma \ref{lemma:beautifyMGF-U}}\label{sec:lemma:beautifyMGF-U}
\begin{lemmaproof}{\ref{lemma:beautifyMGF-U}}
The proof follows after some elementary calculations. That is,
\begin{eqnarray} 
\lefteqn{
 \sum_{i=0}^{ (\ln n)^2} i \  \mathbb{E} _{{\cal F},\MaxDegEventB } [W^t(w_j)|   \OutPathDeg(w_j)=i]
\ \textrm{P}_{{\cal F},\MaxDegEventB }   [\OutPathDeg(w_j)=i ]}
\hspace{.2in} \nonumber \\
&\leq &\left(1+\gamma\right)^{-t}\sum_{i=0}^{(1+\alpha)d} 
i \ \textrm{P}_{{\cal F},\MaxDegEventB }  [\OutPathDeg(w_j)=i ]+
d^{ct}\sum_{i=(1+\alpha)d-2}^{ (\ln n)^2}(i+2)^{t+1} \  \textrm{P}_{{\cal F},\MaxDegEventB }   [\OutPathDeg(w_j)=i  ] 
\nonumber\\
&\leq &\frac{(1+\gamma)^{-t}}{\Pr[ \MaxDegEventB \ | \ {\cal F}_{L}]}
\sum_{i=0}^{(1+\alpha)d}j \Pr[\OutPathDeg(w_j)=i\  | \  {\cal F}_{L}] +
\frac{d^{tc}}{\Pr[ \MaxDegEventB \  |\  {\cal F}_L]}\sum_{i=(1+\alpha)d-2}^{(\ln n)^2}(i+2)^{t+1}\ \Pr[\OutPathDeg(w_j)=i \ |\  {\cal F}_L] \nonumber\\
&\leq &2{(1+\gamma)^{-t}}\sum_{i=0}^{n}j \Pr[{\cal B}(n,d/n)=i]+
2{d^{tc}}\sum_{i=(1+\alpha)d-2}^{(\ln n)^2}(i+2)^{t+1}\ \Pr[\OutPathDeg(w_j)=i \  | \ {\cal F}_L] \nonumber
\end{eqnarray}
In the last derivation we used the fact that $\OutPathDeg(w_j)$ is dominated by ${\cal B}(n,d/n)$ and 
$\Pr[\MaxDegEventB\ | \ {\cal F}_L]\geq 1/2$. It is direct that the first summation is equal to $d$.
As far as the second summation is regarded we use the Claim \ref{claim:WtBoundTech} 
(the claim appears in  the proof of Lemma \ref{lemma:BoundEexpRv}, Section \ref{sec:prop:BoundEexpRv})

We have that 
\begin{eqnarray} 
\lefteqn{
 \sum_{i=0}^{ (\ln n)^2} i  \   \mathbb{E} _{\MaxDegEventB, {\cal F}} [W^t(w_j) \ |\    \OutPathDeg(w_j)=i]
\ \textrm{P}_{{\cal F},\MaxDegEventB }   [\OutPathDeg(w_j)=i ]}
\hspace{3in} \nonumber \\
 &\leq& 2d{(1+\gamma)^{-t}}+2{d^{t(1+c)+1}}\exp\left(-\alpha^2d/5\right). \nonumber\\
  &\leq& 2d{(1+\gamma)^{-t}}+ \exp\left(-\alpha^2d/6\right), \nonumber
\end{eqnarray}
where in the last derivation we used the fact that $t\leq \gamma^{-3}\ d^{7/10}$.
The lemma follows.
\end{lemmaproof}

\section{Remaining Proofs}

\subsection{Proof of Corollary \ref{cor:UpdateComplexity}}\label{sec:cor:UpdateComplexity}

Recall that we have assumed that each $B\in {\cal B}$ is a tree with at most one extra edge.
For such simple structured blocks one may find many different ways of implementing the transitions 
of the chain efficiently. In what follows we describe one possible approach.

Let $(X_t)_{t\geq 0}$ and $(Y_t)_{t\geq 0}$ be the block dynamics for the colouring and 
the hard-core model, respectively.
For both $(X_t)_{t\geq 0}$ and $(Y_t)_{t\geq 0}$,  the update of the configuration of a block
$B\in {\cal B}$ is done assuming a predefined order of the vertices, e.g., $w_1, w_2, \ldots, w_{\ell}$, 
where $\ell=|B|$.  The update of $B$ is done  by assigning  sequentially spins to the vertices. 
That is, having fixed  the spin of the vertices up to some vertex $w_{i}$,  we assign
spin to the next vertex $w_{i+1}$ by working as follows: 
We compute the  distribution of the spin  on the vertex $w_{i+1}$ given the configuration of 
$w_1, \ldots, w_{i}$  and that of $V\setminus B$.
Once we have this distribution, we can assign spin to the vertex $w_{i+1}$ appropriately. 
The critical issue is how do we compute this distribution. Depending on the model 
we follow a different approach.

For $(X_t)_{t\geq 0}$  it suffices to count the number  of $k$-colouring of $B$ which assign $v$ colour $c$, for every $c\in [k]$.
So as to achieve this, we use the Dynamic Programming algorithm for counting colourings 
suggested in \cite{old-GnpSampling} (See the algorithm in Section 3.3 of \cite{old-GnpSampling}).
For fixed $k$ this algorithm is polynomial in $n$.

For $(Y_t)_{t\geq 0}$   it suffices to compute $p_{occ}(w_{i+1})$ the  probability of the vertex $w_{i+1}$ to 
be ``occupied'' (to be in the independent set).   We can use the algorithm by D. Weitz in \cite{Weitz} to compute this probability. 
For  the kind of blocks we consider here, this algorithm  computes exactly $p_{occ}(w_i)$ in 
polynomial time as the size of the so-called "tree of self avoiding walks"  of $B$ has size $O(n)$.

\subsection{Proof of Lemma \ref{lemma:SimpleBlock}}\label{sec:lemma:SimpleBlock}

Let ${\cal B}_1$ be the set of blocks created from the cycles in ${\cal C}$ and let ${\cal B}_2={\cal B}\backslash {\cal B}_1$.
It suffices to show that with probability at least $1-2n^{-3/4}$, ${\cal B}_1$ contains only unicyclic blocks and  ${\cal B}_2$
contains only trees.

First we focus on  ${\cal B}_1$. It suffices to show that with sufficiently large  probability $\G$  is such that no two cycles
in ${\cal C}$ are close to each other,  e.g., at distance smaller than $10\frac{\ln n}{ (\ln  d)^5}$.  
Then, Theorem \ref{thrm:BreakablePaths}  guarantees that the no two cycles will get to the same block.

If there is a pair of cycles in ${\cal C}$ at distance less than  $10\frac{\ln n}{ (\ln  d)^5}$, then the following should hold: There 
is a set of vertices $S$ of cardinality less than $2\frac{\ln n}{(\ln  d)^2}$ such that the number of edges between the vertices 
in $S$ is at least $|S|+1$.  We  show that such a set does not exist in $\G$  with probability at least $1-n^{-3/4}$.

Let $D$ be the event that such a set exists. It holds that
\begin{eqnarray}
\Pr[D]&\leq &\sum_{r=1}^{2\frac{\ln n}{ (\ln  d)^2}}{n \choose r} {{r\choose 2} \choose r+1}\left(\frac{d}{n}\right)^{r+1}
\leq \sum_{r=1}^{2\frac{\ln n}{ (\ln  d)^2}} \left(\frac{ne}{r}\right)^r\left(\frac{r^2e}{2(r+1)}\right)^{r+1}\left(\frac{d}{n}\right)^{r+1}
\hspace{.75cm} \left[\textrm{as }{n \choose r}\leq \left(\frac{ne}{r}\right) \right] \nonumber\\
&\leq& \frac{1}{n}\sum_{r=1}^{2\frac{\ln n}{ (\ln  d)^2}}\left(\frac{erd}{2}\right)\left(\frac{e^2d}{2}\right)^{r}
\leq \frac{ed}{ (\ln  d)^2}\ \frac{\ln n}{n}\sum_{r=1}^{2\frac{\ln n}{ (\ln  d)^2}}\left(\frac{e^2d}{2}\right)^{r}
\hspace{3.15cm}\mbox{[as $r\leq 2\ln n/ (\ln  d)^2$]}\nonumber \\
&\leq& n^{-9/10}\left({e^2d}/{2}\right)^{2\frac{\ln n}{ (\ln  d)^2}}
\leq n^{-3/4}, \nonumber
\end{eqnarray}
where the last two inequalities hold for large fixed $d$ and large $n$. The above proves the part for ${\cal B}_1$, i.e.
\[
\Pr[{\cal B}_1 \ \textrm{contains only unicyclic block}]\geq 1-n^{-3/4}.
\]

\noindent
So as to show that ${\cal B}_2$ consists of tree-like blocks we work as follows: Let some $B\in {\cal B}_2$ and let $w$ be
the vertex we used to created it.  It is direct that every path that connects $w$ to some vertex in any of the blocks in ${\cal B}_1$
should contain at least one break-point (otherwise $w$ should belong to a block in ${\cal B}_1$).  
That is, if $B$ contains a cycle $C$,  then $C\notin {\cal C}$. This implies that $|C|> 4\frac{\ln n}{ (\ln  d)^5}$.  
It suffices to show that with 
probability at least $1-n^{-3/4}$,  every $B\in {\cal B}_2$ cannot contain a cycle of length $\ell\geq 3\frac{\ln n}{ (\ln  d)^5}$.

From Theorem \ref{thrm:BreakablePaths} we  have that   the set   $\mathbf{U}$ of the  { elementary} paths in $\G$ of  
length  $\frac{\ln n}{ (\ln d)^{5} }$  that do not have any break-point  is empty with probability $1-o(1)$.
If  $\mathbf{U}=\emptyset$, every vertex in $B$ should be within distance at most $r=\frac{\ln n}{ (\ln d)^5}$
from $w$. 
We will show that with sufficiently large probability we have  $\mathbf{U}=\emptyset$ and $G\left(w,r \right)$, the induced 
subgraph of $\G$ that contains $w$  and all the vertices within graph distance $r$, is either a tree or unicyclic.
Then,  if there is a cycle in $G\left(w,r \right)$ its length should not exceed  $2\frac{\ln n}{ (\ln d)^5}+1<
3\frac{\ln n}{(\ln  d)^5}$. 
Summarizing,  the above argument implies the that
\[
 \Pr[{\cal B}_2 \ \textrm{contains non-trees}] \leq 
 \Pr[\mathbf{U}\neq \emptyset]+ \Pr[\exists w\ s.t. G\left(w,r \right) \textrm{has more than 1 cycles}].
\]

\noindent
We show that   there is a vertex $w$ in $\G$ such  $G\left(w,r  \right)$ contains more than one cycle with probability 
at most $n^{-3/4}$. To do this we work as we did for the ${\cal B}_1$.
Additionally, we have  $\mathbf{U} \neq \emptyset$ with  probability less than $n^{-2}$, from Theorem \ref{thrm:BreakablePaths}.
With these arguments, the above inequality for ${\cal B}_2$ writes as follows.
\[
\Pr[{\cal B}_2 \ \textrm{contains non-trees}] \leq  2n^{-3/4}.
\]
The lemma follows.

\subsection{Proof of Lemma \ref{lemma:FastBlockCreation}}\label{sec:lemma:FastBlockCreation}

Once the algorithm has the set of break points of $\G$, typically, the construction of
the blocks can be implemented efficiently. This follows from the observation that regardless of 
starting from a cycle or a vertex, so as to recover the block we have to a explore a small, very 
simple-structured neighbourhood. In particular, this exploration can be done efficiently once 
$\G$ has the following properties: $\mathbf{U}=\emptyset$, the cycles in ${\cal C}$ are far parts, 
i.e. at distance greater than $10\frac{\ln n}{(\ln d)^5}$ and the radius $r=\frac{\ln n}{(\ln d)^5}$
neighbourhood  of each vertex in $\G$ is a tree with at most one extra edge. 
As we saw in the proof of  Lemma \ref{lemma:SimpleBlock}, $\G$ has the above properties with
probability greater than $1-10n^{-3/4}$.

The lemma will follow by showing that with probability at least $1-n^{-3/4}$ over the instances of $\G$,  
we can distinguish whether some vertex is break point or not in polynomial time.

For a specific vertex $v$ we need to check the weight of all  paths that start from $v$ and are of length at most $\frac{\ln n}{d^{2/5}}$, 
i.e. we have to check all the paths in ${\cal P}(v)$.
Working as in the proof of Lemma \ref{lemma:SimpleBlock} we have that with probability at least $1-n^{-3/4}$ for every vertex 
$v$ in  $\G$ the neighbourhood we need to check is   a tree with at most one extra edge.  
That is, there are at most 2 different paths between $v$ and a vertex $u$ at distance at most $\frac{\ln n}{d^{2/5}}$. 
The number of paths we need to consider is trivially  upper bounded by $2n$,
while  the computation of the weight  of a specific path $L$ requires $O(|L|)$  elementary arithmetic operations, i.e. $O(\ln n)$ operations. 

We conclude that indeed with probability at least $1-n^{-3/4}$ over the instances $\G$,  for each vertex $v\in V(\G)$ 
the number of steps required to decide weather this vertex  is a break-point or not requires $O(n \ln n)$ steps. 
The lemma follows.

\subsection{Proof of Lemma \ref{lemma:ExpDisGraphCond}}\label{sec:lemma:ExpDisGraphCond}

As far as (a) is regarded, we use the result in \cite{k-core}, which states that w.h.p. $\G$ has no $t$-core for $t\geq d$.
As far as (b) is regarded we use the result from \cite{optas-chrno,chromatic2},  i.e. with probability $1-o(1)$ the chromatic number of $G(n,d/n)$
is $d/(2\ln d)$. As far as (c) is regarded we use Lemma \ref{lemma:SimpleBlock}  from Section \ref{sec:BlckCreation}.

As far as (d) is regarded observe the following:  There is no break-point at the outer boundary of a block $B$  that is influenced 
by a vertex within distance $\ln n/d^{2/5}$  inside $B$.  It remains to consider the paths of length larger than $\ln n/d^{2/5}$.

From Lemma \ref{lemma:SimpleBlock} we have that every $B\in {\cal B}$ is a tree with at most one extra edge. Furthermore, 
the fact that  ${\cal C}$ contains only cycles of length less than $10\log n/(\log d)^2$ and the fact that
there is no elementary path of length greater than $\log(n)/(\log d)^2$, imply that in each $B\in {\cal B}$ there is
no path of length greater than $50\ln n$.

Let $Z_i$ be the number of paths $L$ in $\G$ such that $|L|=i$ and  $\prod_{u\in L}W(u)>1$.
Also, let $\rho_i$ be the probability for  $L$ such that $|L|=i$ to have $\prod_{u\in L}W(u)>1$. 
Using  Theorem \ref{thrm:CL-Tail} 
we have that
\begin{equation}
\rho_i\leq \exp\left(-d^{4/5} i \right), \qquad \forall i\leq 100\ln n. \label{eq:rho_iBound}
\end{equation}
Let $Z=\sum^{50\ln n}_{i\geq {\ln n}/{d^{2/5}}}Z_i$. It is direct to see that if $Z=0$, then the condition (d) holds. 

Let $\ell=\frac{\ln n}{d^{2/5}}$. From Markov's  inequality we have that
\begin{eqnarray}
\Pr[Z>0]&\leq& \mathbb{E}[Z]\leq 
\sum_{i \geq  \ell} n^{i}\left(\frac{d}{n}\right)^{i-1}\rho_i\leq (n/d)\sum^{50\ln n}_{i\geq \ell } d^{i} \ \rho_i
\nonumber \\
&\leq& n\sum^{50\ln n}_{i\geq \ell } \exp\left(-i(d^{4/5}-\ln d)\right) 
\qquad\qquad\qquad \qquad \mbox{[from \ref{eq:rho_iBound}]}
\nonumber\\
&\leq& n^{-\frac{d^{2/5}}{2}}\sum_{i=0}^{\infty} \exp\left(-{d^{4/5}i}/{2} \right)
\leq 2n^{-\frac{d^{2/5}}{2}}. \nonumber
\end{eqnarray}
The above implies that   (d) is satisfied with probability at least $1-2n^{-\frac{d^{2/5}}{2}}$ over the instances of $\G$. 

The lemma follows.

\subsection{Proof of Lemma \ref{lemma:RvsRBreaks}}\label{sec:lemma:RvsRBreaks}

Assume a process where we check which of the vertices in $L=w_1,\ldots,w_{\ell}$ are left breaks, starting from 
$w_1$, then $w_2$ and so on. Since we are interested in left breaks, so as to check $w_{j€™}$ we don't have to examine 
the influence of paths getting to this  vertex from its neighbor $w_{j€™+1}$. 

Assume that the process is to check $w_i\in L$. Let $w_j$, with $j<i$ be the last left break the process has encountered.
If it hasn't found any left break  yet, we set $j=0$. For the lemma, it suffices to show that  $\prod_{r=j+1}^iU(w_r)$ is an 
upper bound on  the influence on $w_i$ from paths that  reach the vertex either from $w_{i-1}$ of from the neighbours of $w_i$ 
outside $L$, i.e. the vertices in $N_i$. We show this by   using induction on the difference $i-j$.

The basis of induction is the case where $i-j=1$.  First we consider the case  where  $j>0$. Then we consider the case 
where $j=0$.

When  $i-j=1$ and $j>0$,  we have that the path arriving either from $w_{j-1}$ or $N_j$ that has   the maximum influence on $w_j$ 
is at most $1$. That is the maximum influence of paths reaching to $w_i$ from $w_j$ is $W(w_i)$, where $W(w_i)$ is  defined in 
(\ref{eq:DefOfW(vi)}). Also, the maximum influence on $w_i$ from paths that pass through its neighbours in $N_i$ is at most 
$\max\{1,Q(w_i)\}\  W(w_i)$. Clearly among the aforementioned paths no path has influence bigger than 
$U(w_i)=\max\{1,Q(w_i)\}W(w_i)$.

When $i-j=1$ and  $j=0$, there are no paths reaching to $v_i$ from $w_j$, i.e.  we have the paths reaching from the 
vertices in $N_i$. Then, it is direct to check that the maximum influence on $w_i$ is at most  $U(w_i)$.

Assume that the  hypothesis holds for $i-j=j_0$, for some $j_0\geq 1$. We are going to show that it also holds for
$i-j=j_0+1$. 

By the  induction hypothesis we have that for $w_{j+j_0}$ the maximum influence from a path that is reaching it
either from $w_{j+j_0-1}$ or from $N_{j+j_0}$ is at most $\prod_{r=j+1}^{j+j_0}U(w_r)>1$. For $i$ such that $i-j=j_0+1$, 
the following  holds: the influence of any path that reaches $w_i$ through $w_{i-1}$ \footnote{ $w_{i-1}$ and  $w_{j+j_0}$ are 
identical}  cannot be larger than $W(w_i)\prod_{r=j+1}^{j+j_0}U(w_r)$ (due to induction hypothesis). Also, there is no path 
reaching $w_i$ from vertices in $N_i$  that has influence larger than $U(w_i)$. It is a matter of direct calculations to
verify that none of the paths we consider has influence greater than $\prod_{r=j+1}^{j+j_0+1}U(w_r)$. 
The lemma follows.

\vspace{.5cm}
\noindent
{\bf Acknowledgement.} The author of this work would like to thank Amin Coja-Oghlan and Eric Vigoda,
for our  discussions and their  comments on this work.

\end{document}